\documentclass[epj]{svjour}
\usepackage{amssymb}
\usepackage{graphics}
\begin{document}
\title{Production of $\mathrm{\Lambda}$ and $\mathrm{\Sigma^0}$ hyperons in proton-proton collisions}

\subtitle{{\it The COSY-TOF collaboration}}

\mail{\mbox{m.schulte-wissermann@physik.tu-dresden.de}  (M.~Schulte-Wissermann)}

\author{M.~Abdel-Bary\inst{3}   \and
S.~Abdel-Samad\inst{3}   \and
K-Th.~Brinkmann\inst{1}$^{,}$ \inst{8}   \and
H.~Clement\inst{4}   \and
J.~Dietrich\inst{1}   \and
E.~Doroshkevich\inst{4}   \and
S.~Dshemuchadse\inst{1}   \and
K.~Ehrhardt\inst{4}   \and
A.~Erhardt\inst{4}   \and
W.~Eyrich\inst{2}   \and
D.~Filges\inst{3}   \and
A.~Filippi\inst{7}   \and
H.~Freiesleben\inst{1}   \and
M.~Fritsch\inst{2}   \and
W.~Gast\inst{3}   \and
J.~Georgi\inst{2}   \and
A.~Gillitzer\inst{3}   \and
J.~Gottwald\inst{1}   \and
D.~Hesselbarth\inst{3}   \and
H.~J\"ager\inst{3}   \and
B.~Jakob\inst{1}   \and
R.~J\"akel\inst{1}   \and
L.~Karsch\inst{1}   \and
K.~Kilian\inst{3}   \and
H.~Koch\inst{9}   \and
M.~Krapp\inst{2}   \and
J.~Kre\ss\inst{4}   \and
E.~Kuhlmann\inst{1}   \and
A.~Lehmann\inst{2}   \and
S.~Marcello\inst{7}   \and
S.~Marwinski\inst{3}   \and
S.~Mauro\inst{9}   \and
W.~Meyer\inst{9}   \and
P.~Michel\inst{5}   \and
K.~M\"oller\inst{5}   \and
H.~P.~Morsch\inst{3}$^{,}$\inst{6}   \and
H.~M\"ortel\inst{2}   \and
L.~Naumann\inst{5}   \and
N.~Paul\inst{3}   \and
L.~Pinna\inst{2}   \and
C.~Pizzolotto\inst{2}   \and
Ch.~Plettner\inst{1}   \and
S.~Reimann\inst{1}   \and
M.~Richter\inst{1}   \and
J.~Ritman\inst{3}   \and
E.~Roderburg\inst{3}   \and
A.~Schamlott\inst{5}   \and
P.~Sch\"onmeier\inst{1}   \and
W.~Schroeder\inst{2}$^{,}$\inst{3} \and
M.~Schulte-Wissermann\inst{1}   \and
T.~Sefzick\inst{3}   \and
F.~Stinzig\inst{2}   \and
M.~Steinke\inst{9}   \and
G.~Y.~Sun\inst{1}   \and
A.~Teufel\inst{2}   \and
W.~Ullrich\inst{1}   \and
G.~J.~Wagner\inst{4}   \and
M.~Wagner\inst{2}   \and
R.~Wenzel\inst{1}   \and
A.~Wilms\inst{9}   \and
P.~Wintz\inst{3}   \and
S.~Wirth\inst{2}   \and
P.~W\"ustner\inst{3}   \and 
P.~Zupranski\inst{6}}

%

%
\institute{Institut f\"ur Kern- und Teilchenphysik, Technische Universit\"at Dresden, D-01062 Dresden, Germany \and
		   Physikalisches Institut, Universit\"at Erlangen-N\"urnberg, D-91058 Erlangen, Germany \and
		   Institut f\"ur Kernphysik, Forschungszentrum J\"ulich, D-52425 J\"ulich, Germany \and
		   Physikalisches Institut, Universit\"at T\"ubingen, D-72076 T\"ubingen, Germany \and
		   Institut f\"ur Kern- und Hadronenphysik, Forschungszentrum Dresden-Rossendorf, D-01314 Dresden, Germany \and
		   Soltan Institute for Nuclear Studies, 05-400 Swierk/Otwock, Poland \and
		   INFN Torino, 10125 Torino, Italy \and
		   Helmholtz Institut f\"ur Strahlen- und Kernphysik, Rheinische Friedrich-Wilhelm-Universit\"at Bonn, D-53115 Bonn, Germany \and
		   Institut f\"ur Experimentalphysik, Ruhr-Universit\"at Bochum, D-44780 Bochum, Germany}
\date{Received: date / Revised version: date}
%

\abstract{ 
This paper reports results on simultaneous measurements of the reaction channels $\mathrm{pp\to pK^+\Lambda}$ 
and $\mathrm{pp\to pK^+\Sigma^0}$ at excess energies of
204, 239, and 284 MeV ($\mathrm{\Lambda}$) and 127, 162, and 207 MeV ($\mathrm{\Sigma^0}$). 
Total and differential cross sections are given
for both reactions. It is concluded from the measured total cross sections that the high energy 
limit of the cross section ratio is almost reached at an excess energy of only about 200 MeV. 
From the differential distributions observed in the overall CMS as well as in the  Jackson and helicity 
frames, a significant contribution of interfering nucleon resonances to the $\mathrm{\Lambda}$ production 
mechanism is concluded while resonant $\mathrm{\Sigma^0}$-production seems to be of lesser importance
and takes place only through specific partial waves of the entrance channel.
The data also indicate that kaon exchange plays a minor role in the case of $\mathrm{\Lambda}$- but an 
important role for $\mathrm{\Sigma^0}$-production. Thus the peculiar energy dependence of the $\mathrm{\Lambda}$/$\mathrm{\Sigma^0}$ 
cross section ratio appears in a new light as its explanation requires more than mere differences 
between the p$\mathrm{\Lambda}$ and the p$\mathrm{\Sigma^0}$ final state interaction. The data provide a benchmark 
for theoretical models already available or yet to come.
}

\maketitle

\section{Introduction}
\label{intro}
The physics program carried out at the COoler SYn\-chro\-tron COSY (Forschungszentrum J\"ulich, Germany)  
focuses on the study of mesons and baryons in the confinement regime of QCD.
As\-so\-ciated strange\-ness production plays a major role within this general field and has been studied 
by various experimental groups at COSY. In the case of proton-proton induced $\mathrm{\Lambda}$ and $\mathrm{\Sigma^0}$ hyperon production this effort has led to very
well measured excitation functions starting at excess energies 
($\epsilon=\sqrt s-(m_p+m_K+m_Y)$) only a few MeV above the thresholds and extending to 
$\epsilon \approx 200\,\mathrm{MeV}$  in the case of $\mathrm{\Lambda}$ and $\epsilon \approx 120\,\mathrm{MeV}$ 
in the case of $\mathrm{\Sigma^0}$ production  \cite{grzonka97,balewski97,balewski98,sewerin99,COSY11kowina04,toflambda06,valdau07}.

One surprising result of these investigations is that the excitation function of $\mathrm{pp\to pK^+\Sigma^0}$ 
exhibits within uncertainty a pure phase space behavior  
($\sigma_{\mathrm\Sigma^0} \propto \epsilon^2$) whereas 
the $\mathrm{pK^+\Lambda}$ final state is produced with larger abundance 
towards the production threshold. This leads to a peculiar energy dependence  of
the production cross section ratio $R_{\mathrm{\Lambda/\Sigma^0}}=\sigma_{\mathrm{\Lambda}} / \sigma_{\mathrm{\Sigma^0}}$.
A few MeV above the threshold a high value of  $R_{\mathrm{\Lambda/\Sigma^0}}=28^{+6}_{-9}$ has been found \cite{balewski98,sewerin99}.
With increasing  excess energy $R_{\mathrm{\Lambda/\Sigma^0}}$ drops steadily to a value of $8\pm2.7$ 
at $\epsilon=60\,\mathrm{MeV}$  \cite{COSY11kowina04}. Then, up to $\epsilon$ = 700 MeV the energy dependence 
of $R_{\mathrm{\Lambda/\Sigma^0}}$ cannot be determined in a model independent manner, as the few existing data points 
were taken at different excess energies.
However, above 700 MeV the data collected in \cite{baldini88} show  $R_{\mathrm{\Lambda/\Sigma^0}}$ to stay essentially
constant at a value of only $2.2\pm 0.2$.
The low and the high energy limits, respectively, are both close to ``magic values'', as a ratio of 27
is the SU(6) prediction of the ratio of the squares of the coupling constants 
of the virtual exchange-kaon ($g^{2}_{N\Lambda K}/g^{2}_{N\Sigma K}$) \cite{deswat63} 
and the ratio of three follows from isospin considerations. 
It has been argued, however, that these theoretical explanations are oversimplifications \cite{sibi06,shyam06}; 
perhaps one observes only a fortuitous coincidence.

The dramatic change of the $\mathrm{\Lambda}$ to $\mathrm\Sigma^0$ cross section ratio with 
excess energy 
could be due to different 
production processes, or distinctively different final state interactions (FSI), or both.
However, the reaction mechanisms are far from being established as
various theoretical approaches have been developed in the last years 
which all reproduce the experimental data with about the same amount of success.
These models  are based on very different footings:
 a coherent interplay of FSI and nucleon resonances was concluded for $\mathrm{\Lambda}$ production \cite{sibi06}; 
 the production via established nucleon resonances was studied \cite{shyam06}; 
 a contribution of an N(1535) resonance without any proton-hyperon FSI was suggested \cite{luizou06};
 destructive interferences between $\mathrm\pi$ and K exchange contributions were considered \cite{gaspa00};
 and a constituent quark-gluon model with the inclusion of nucleon resonances was proposed \cite{dillig06}.
Finally, a model based on kaon and pion exchange developed for SATURNE \cite{saturneandfriends1,saturneandfriends2} 
data ($\epsilon>1\,\mathrm{GeV}$) long before the COSY era
also reproduces the energy dependence of $R_{\mathrm{\Lambda/\Sigma^0}}$ fairly well, as
without any adjustment of the parameters the ratio is only underestimated by 50\% directly at threshold  \cite{laget01Nlaget91}.
Hence,
the energy dependence of the total cross sections alone 
is not at all sufficient to discriminate between these different theoretical approaches.
Differential observables for both reaction channels are highly desired in order to set benchmarks 
for theoretical models.

\begin{figure}[ttt]
\resizebox{.5\textwidth}{!}{%
 \includegraphics{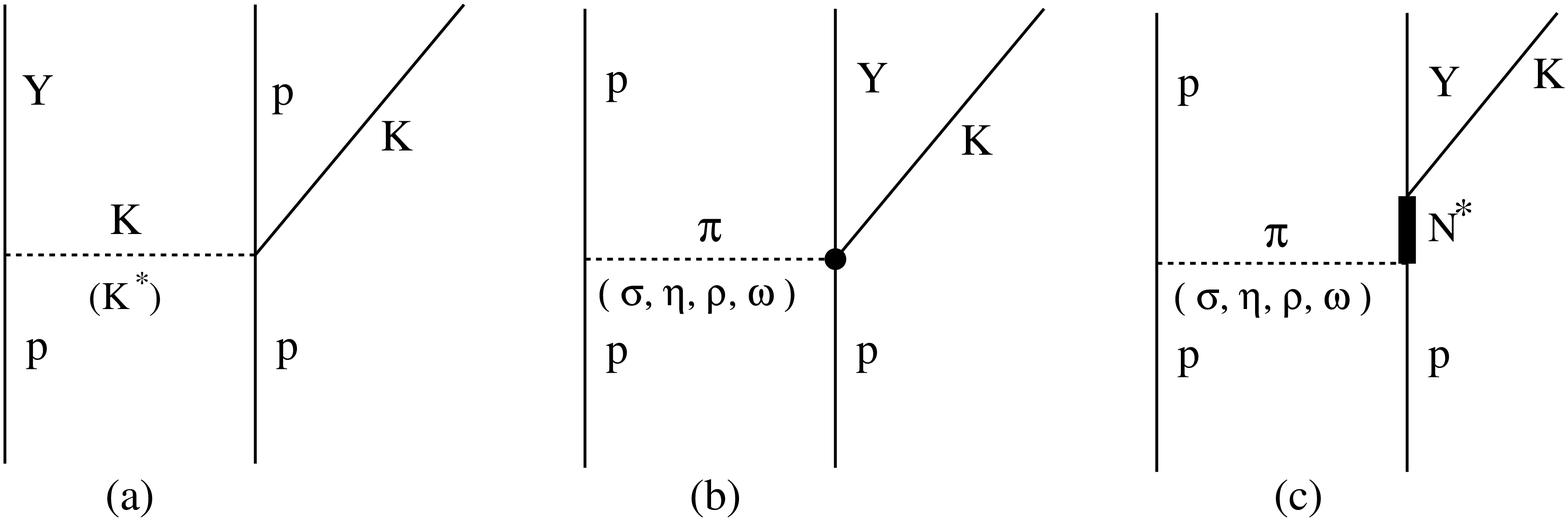} }
\caption{Reaction mechanisms involving (a) strange, (b) non-resonant, and (c) resonant, non-strange meson exchange 
for $\mathrm{pp\to pK^+}Y, Y=\mathrm{\Lambda,\Sigma^0}$. 
Initial and final state interactions are not indicated.}
\label{fig:ractionmechnism}  
\end{figure}

This large variety of theoretical approaches to describe hyperon ($=Y$) production 
in proton-proton collisions can be 
or\-dered into two classes of pro\-duction scenarios depicted in 
fig.~\ref{fig:ractionmechnism}. On the one hand, 
kaon (and $\mathrm{K^*}$) exchange manifests strangeness in the production mechanism itself. 
In this case resonances could be involved in principle, however, none is known in the Kp-system at present.
On the other hand, 
$\mathrm{\pi}$ (and $\mathrm{\sigma}$, $\mathrm{\eta}$, $\mathrm{\rho}$, $\mathrm{\omega}$) 
exchange shifts the strangeness production away from the interaction of both protons
to the p$\pi\to KY$ vertex.
Since the subprocess p$\pi \to KY$ is likely to involve an intermediate resonance 
(p$\pi \to N^*,\Delta^*; N^*/\Delta^* \to KY$), 
this scenario suggests the role of resonances to be of importance for proton-proton induced hyperon production.
 
Numerous studies of nucleon and $\Delta$ resonances excited in $\mathrm{\pi N}$ and $\mathrm{\gamma N}$
reactions revealed among other resonances those which decay into particles
with open strangeness such as the associated pairs $\mathrm{K\Lambda}$ or $\mathrm{K\Sigma}$
\cite{capstickNroberts98,bradford06,sumihama06,hicks07,mart06,tram98}.
According to these findings and other experimental as well as theoretical work, the particle data group
\cite{pdg06} lists the resonances $\mathrm{N(1650)S}_{11}$,
$\mathrm{N(1675)D}_{15}$, $\mathrm{N(1680)F}_{15}$, $\mathrm{N(1700)D}_{13}$,
$\mathrm{N(1710)P}_{11}$, $\mathrm{N(1720)P}_{13}$, and $\mathrm{N(1900)P}_{13}$ and values
for the $\mathrm{K^+\Lambda}$ branching ratios.
Strong evidence was advanced for additional states, namely $\mathrm{N(1840)P}_{11}$
and $\mathrm{N(1875)D}_{13}$ \cite{sarantsev05}. The latter may be the same
as $\mathrm{N(1895)D}_{13}$ \cite{McNabb04,glander03} or $\mathrm{N(1950)D}_{13}$ \cite{pennerNmosel02}.
The properties of all these resonances (mass, width, branching ratios to K$\mathrm{\Lambda}$ or K$\mathrm{\Sigma}$) are typically not known
well and
the role these nucleon resonances play for associated strangeness production in proton-proton collisions
is not well understood.
It should be mentioned that the concept of resonant hyperon production limits
the $\mathrm{K^+\Lambda}$ final state to be produced only from $N^*$ resonances (due to isospin conservation)
whereas the $\mathrm{K^+\Sigma^0}$ pair can be reached via both $N^*$ and $\Delta^*$ resonances. In the latter case,
$\mathrm{\Delta(1600)P_{33}}$, $\mathrm{\Delta(1620)S_{31}}$,
$\mathrm{\Delta(1700)P_{33}}$, $\mathrm{\Delta(1750)P_{33}}$, and $\mathrm{\Delta(1900)S_{31}}$ can also be involved.
The $\Delta^*$-resonances increases the number of possible contributions to the $\mathrm{K^+\Sigma^0}$ production process
and renders a theoretical description even more difficult than that of the $\mathrm{pK^+\Lambda}$ channel.
Turning the argument around the isospin-selective $\mathrm{pp\to pK^+\Lambda}$ channel could potentially serve as a tool
to find so-called missing resonances (with I=1/2) \cite{capstickNroberts2000}, i.e.~resonant states predicted
by quark-models which have not been found experimentally so far.

The {\it COSY-TOF} collaboration has recently published data on total cross sections 
for the $\mathrm{pp\to pK^{+}\Lambda}$ channel at moderate excess energies
of 85, 115, and 171 MeV, and the first differential data, namely  Dalitz plots and 
helicity angle distributions at $\epsilon$=171 MeV \cite{toflambda06}. 
In addition, invariant mass spectra are shown.
The results can be described  by a reaction model which includes 
in a coherent manner the $\mathrm{N(1650)S}_{11}$, $\mathrm{N(1710)P}_{11}$, 
and $\mathrm{N(1720)P}_{13}$ nucleon resonances 
in conjunction with a considerable contribution of $\mathrm{p\Lambda}$ final-state-in\-ter\-action.
With increasing energy the relative contribution of the $\mathrm{N(1650)S}_{11}$ then diminishes in trade for a stronger
influence of the $\mathrm{N(1710)P}_{11}$ and $\mathrm{N(1720)P}_{13}$ as shown in  \cite{eyrich07,schoeder07}.

The only known experimental fact in the case of the $\mathrm{pp\to pK^{+}\Sigma^0}$ reaction is
the proportionality of the total cross section to the phase space volume ($\sigma_{\mathrm{\Sigma^0}} \propto \epsilon^2$).
This can be taken as a indication for the reaction mechanism to be independent of the excess energy
and the absence of a strong $\mathrm{p\Sigma^0}$-FSI which would manifest itself close to threshold.
However, it is of course possible that competing processes cancel in a way that they mimic 
the $\sigma_{\mathrm{\Sigma^0}} \propto \epsilon^2$ behavior.
As experimental results on differential observables are completely lacking for $\mathrm{\Sigma^0}$ production in pp-collisions 
any detailed theoretical approach is currently hampered strongly. 

In this paper differential distributions for both reactions are presented, namely 
the angular distributions of all final state particles in the overall CMS, 
as well as distributions in both the Jackson and helicity frames of all two-body subsystems. 
Like Dalitz plots, the helicity angle distributions 
provide insight into the three-body final state. They are especially well suited to investigate the  
influence of intermediate resonances. 
The information contained in the Jackson angle distributions are complementary to that of a Dalitz plot, as this angular distribution
relates the exit to the entrance channel and hints at relative angular momenta 
and/or resonances present in a specific two-body subsystem. 

The data were taken at beam momenta of $p_{beam} = 2950, 3059$, and $3200\,\mathrm{MeV/c}$. 
These beam momenta corres\-pond to excess energies of 204, 239, and 284 MeV in the case of 
$\mathrm{pp\to pK^+\Lambda}$ ($m_{\mathrm\Lambda}=1116\,\mathrm{MeV}$) while
 the neutral $\mathrm{\Sigma}$  is produced  127, 162, and 207 MeV above 
threshold ($m_{\mathrm\Sigma^0}=1193\,\mathrm{MeV}$).
The data taken at $p_{beam}$ $=3059 \,\mathrm{MeV/c}$ stand out since they have a large in\-tegra\-ted lu\-minosity; 
in this case the calibration, the acceptance\footnote{The term ``acceptance'' is used for the 
convolution of solid angle coverage, detector-, and reconstruction-efficiency.} 
correction as well as the overall luminosity 
were cross-checked by three independent subgroups 
of our collaboration when a supposed pentaquark state was searched for \cite{tofthetaplusdead}. 
The measurements at 2950 MeV/c and 3200 MeV/c  were carried 
out directly one after the other without any change of 
detector set-up, DAQ, high voltage or electronics. 
Hence,  the results of these two data sets are ideal for relative comparisons as 
systematic uncertainties partly cancel.

\begin{figure}[tth]
\resizebox{0.5\textwidth}{!}{ \includegraphics{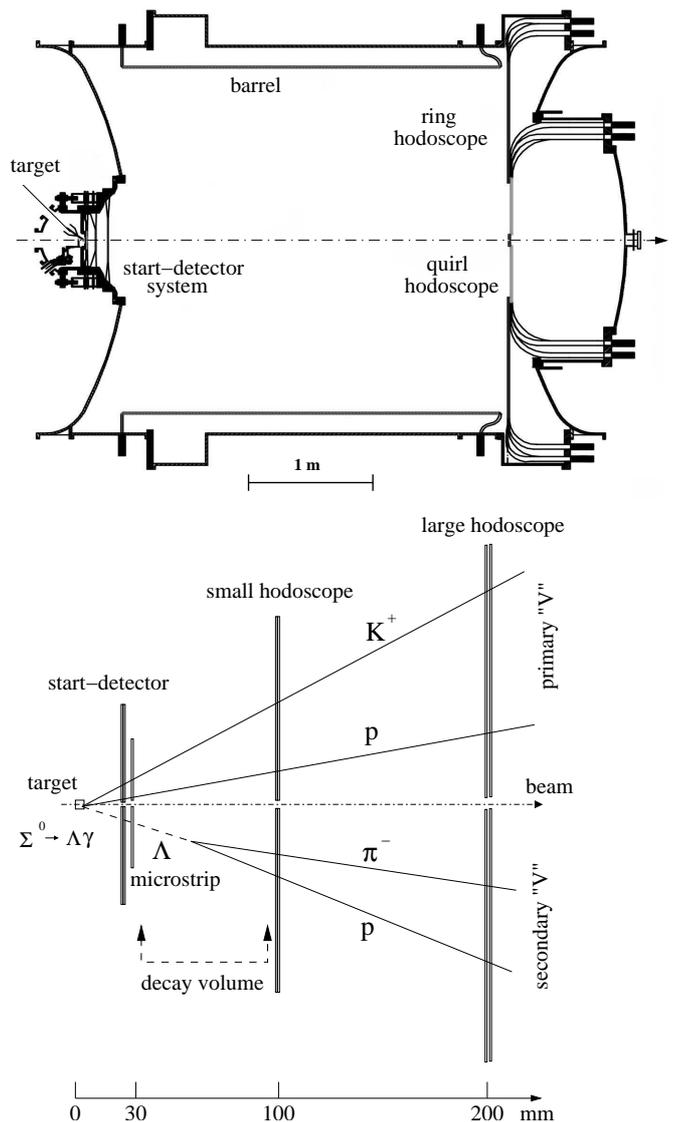} }
\caption{The {\it COSY-TOF} detector (top), the near-target region (start-detector system, bottom).
Within the lower picture the event pattern for both hyperon channels is shown.
}
\label{fig:tofNestart}  
\end{figure}

\section{Experimental procedure}
\subsection{Detector setup}
The experiments were carried out with the time-of-flight detector {\it COSY-TOF} located at an external beam line of 
the COoler SYnchrotron COSY (For\-schungs\-zen\-trum J\"ulich).
The COSY machine provides proton beams of very high quality (spill length $\approx$ 5 min; several $10^6$
protons/s; low emittance of $< 5\, \pi\, \mathrm{mm\, mrad}$; 
relative momentum uncertainty  $\Delta p/p < 10^{-3}$).

The layout of the {\it COSY-TOF} detector is shown in the upper part of fig.~\ref{fig:tofNestart}; 
 in the lower part the near target region with the 
time-of-flight start and tracking detectors \cite{tofbeschreibung,Erlangenstart} is sketched.
The interaction volume is small and well defined as the narrow beam with 
Gaussian profile ($\sigma_{x,y}<300\,\mathrm{\mu m}$) is directed onto a liquid hydrogen target of only 4 mm length \cite{target}.
The emerging particles traverse just behind the target ($\approx 25$ mm) a 24-fold segmented 
scintillation detector (``start-detector'')
which provides the start signal for the time-of-flight measurement. 
At a distance of 30 mm downstream of the target a double-sided silicon-microstrip detector is installed, followed by
two double-layered scintillating fiber hodoscopes at 100 and 200 mm.  
These three tracking detectors measure the coordinates of traversing 
charged particles in three dimensions with a spatial resolution of $\approx 100\,\mu m$ 
(microstrip) and $\approx1.5$ mm (hodoscopes).
 
After a flight path of $\approx\negthinspace$ 3 m through the evacuated ves\-sel (0.2 Pa) all charged particles 
are detected in the highly segmented stop components. They consist of two triple-lay\-ered 
forward hodoscopes (central and ring hodoscope) \cite{forewardhodo}
and a barrel hodo\-scope \cite{barrel}, all manufactured from BC412 scintillating material. 
From the combined measurement of time and position the velocity vectors of all charged particles
are determined with a time-of-flight resolution of better than $\sigma_{TOF}$ = 300 ps 
and an angular track-resolution of better than $\sigma_{\sphericalangle}=0.3^\circ$. 
Primary vertices (located within the target) are reconstructed with an accuracy of better than 
$\sigma_{x,y} = 0.5\,\mathrm{mm}$ and $\sigma_{z} = 2.0\,\mathrm{mm}$.
Secondary vertices from particles
decaying behind the microstrip detector and before the first hodoscope are reconstructed with 
an accuracy of $\sigma_{x,y} < 1\,\mathrm{mm}$ and $\sigma_{z} < 3.0\,\mathrm{mm}$.

The {\it COSY-TOF} detector stands out for its low mass area density of target, start-detector, and tracking detectors. 
This renders
the influence of small angle scattering and energy loss almost negligible.
In addition, the {\it COSY-TOF} detector has a high efficiency of $>95\% 
$ for the detection of charged particles and covers a large solid angle
($1^\circ<\theta<60^\circ,\, 0^\circ<\phi<360^\circ$) in the laboratory frame.
These features allow the almost unambiguous and simultaneous identification of different reaction channels 
({\it e.g.}, $\mathrm{pp\to pp}$, $\mathrm{d\pi^+}$, $\mathrm{pp\omega}$, 
$\mathrm{pK^+\Lambda}$, $\mathrm{pK^+\Sigma^0}$, $\mathrm{pK^0\Sigma^+}$)
by examining the measured time-of-flight of the charged particles and their event topology.

\subsection{Principle of measurement and data analysis}
The strategies presented in this paper for the analysis of neutral hyperon  
production are a straightforward extension of the 
special routines developed by the Dresden group for the analysis 
of the reaction channel $\mathrm{pp\to pK^0\Sigma^+}$ 
(analysis A in \cite{tofthetaplusdead}) within our standard analysis framework 
\cite{MSWdiss,WU07,mswpaper06,wupaper09}.
This approach is an alternative one to that applied in \cite{toflambda06,Erlangenstart}.

Here, the focus lies on an extensive use of the time-of-flight information in order to independently determine, 
on the one hand, 
the primary particles ($\mathrm{p, K^+}$) and, on the other hand, 
the secondary particles stemming from the hyperon's decay into $\mathrm{p}$ and $\mathrm{\pi^-}$. 
While the secondary particles are used as a ``positive tag'' for a $\mathrm{pK^+Y}$ events, the four-momenta 
of primary protons and kaons are utilized to calculate the missing mass spectrum which 
shows both hyperons well resolved. This missing mass spectrum is then the backbone of the analysis.

The simultaneous analysis of the reactions $\mathrm{pp}\to \mathrm{pK^{+}\Lambda}$ and 
$\mathrm{pp\to pK^+\Sigma^0}$ is possible due to the decay properties of the  hyperons involved.
A $\mathrm{\Sigma^0}$ ($m_{\mathrm{\Sigma^0}}=1193\, \mathrm{MeV}/\mathrm{c}^2$) decays with a 
branching ratio 
of $\mathcal{BR}\approx 100\%$ in 
the direct vicinity of its production vertex ($\mathrm{c}\tau =  2.2\cdot 10^{-11} \mathrm{m}$)
to  $\mathrm{\Lambda}\gamma$.
The photon remains undetected, however, its energy is small compared to
the mass of the $\mathrm{\Lambda}$ ($m_{\mathrm{\Lambda}}=1116\,\mathrm{MeV}/\mathrm{c}^2$) 
and thus the change of direction of the $\mathrm{\Lambda}$ 
with respect to that of the $\mathrm{\Sigma^0}$ is less than $5^\circ$ in the laboratory system. 
This minor change of direction justifies to analyze both reactions on equal footing.

During data taking both final states are preselected via the same
trigger based on the mul\-ti\-pli\-city-jump of charged particles (see fig.~\ref{fig:tofNestart}, lower part): 
near the target (start-detector) only two charged particles are found (p, K$^+$)
whereas, due to the sizable decay length of the $\mathrm{\Lambda}$ hyperon ($\mathrm{c}\tau_\Lambda = 7.89\,\mathrm{cm}$) 
and its decay into two
charged particles ($\mathrm{\Lambda \to p\pi^-},\, \mathcal{BR} = 64.2\%$),
four hits are detected in the stop components. 
This requirement is also modeled as the first step in the off-line analysis.

Both primary ($\mathrm{p,K^+}$) and secondary ($\mathrm{p,\pi^-}$) particles emerge from 
a  vertex and form a $V$-shaped pattern in the detector.
These patterns are searched for by a fitting procedure,
where each layer of the tracking detectors, which has produced a signal, 
provides an individual track point. 
In order to identify these two different sets of $V$ shaped patterns  
the following self-evident conditions are exploited:
a primary $V$ is characterized by two hits in the stop components in coincidence with two hits in the 
start-detector together with any number of track points in the microstrip and hodoscopes;
the vertex must be located inside the target within resolution. 
A secondary $V$ has two hits in the stop components; at least 
three track points in each arm; a decay plane intersecting the target 
volume; and a decay vertex 
located in the volume between the microstrip detector and the first hodoscope (decay volume).
All permutations of hits in the stop 
components are taken into account resulting first in a set of primary and then in secondary $V$-candidates. 
The best $V$s are chosen according to the number of track points and the quality of the fit.
This method was developed by means of Monte Carlo data. 

So far only geometric information has been exploited in order to independently determine one 
primary and one secondary $V$.
Using the measured time-of-flight ($t_{stop} - t_{start}$) for the particles of each arm of the primary $V$s 
leads to corresponding primary velocity vectors. 
An averaged start-detector value is used for the time-of-flight determination of secondary particles.
Masses have to be assigned to these velocity vectors in order to obtain four-vectors. 
As the {\it COSY-TOF} detector does not provide direct particle identification, 
the event-topology is used. 
For the secondary $V$, the particle with the smaller angle to the 
direction of flight of the 
hyperon (determined from the direction from the primary to the decay vertex) is called ``proton'' while the other 
is called ``pion''. Monte Carlo studies have shown  that, 
due to the large mass difference of both particles, this assignment is correct for more than 99\% of the $\mathrm{pK^+\Lambda}$ events
(97\% in case of $\mathrm{pK^+\Sigma^0}$).

In order to label each arm of the primary $V$ correctly as proton and kaon the direction of flight of the hyperon is used.
This observable is measured independently twice, 1) from the vector connecting the origin 
with the vertex of the secondary $V$ and
2) from the missing momentum vector calculated from the primary particles. 
The latter is calculated for both possibilities of particle assignment, the one with the better match of flight directions 
is chosen to be the proper one.
Monte Carlo studies have shown that the mass assignment is correct for about 80\% of the events. Swapped mass 
assignments lead to a broad missing mass distribution in the final data sample with no peaks in the vicinity of the hyperon masses.
 
The background is reduced by requiring the properties of the secondary $V$ 
to match those of a decaying $\mathrm{\Lambda}$ hyperon. 
Firstly, it is necessary that the angle of the secondary proton 
with respect 
to the $\mathrm{\Lambda}$ flight direction lies within the 
kinematically possible region ($<10^\circ$). 
Secondly, the invariant mass of proton and pion, $m_{\mathrm{p\pi^-}}=m_\mathrm{\Lambda}$,
is calculated from the flight direction of the hyperon (decay vertex), 
the measured four momentum of the secondary proton, and the direction vector of the pion. 
This value must match the $\Lambda$ mass within limits determined by Monte Carlo.
The combination of both requirements reduces Monte Carlo and data signals  by only 4\% while 
40\% of the experimental background is suppressed in the final data sample.

As a final selection criterion, only events with the combined momentum vector of primary proton and kaon
pointing in the backward CMS hemisphere are considered. 
Due to the Lorentz boost the particles in this hemisphere
have smaller velocities in the laboratory system.  This increases the relative time-of-flight resolution, which in turn significantly
increases the absolute momentum, and hence, the missing-mass resolution.
It should be mentioned that this requirement leads to no loss of physical information, 
as the symmetric entrance channel (proton-proton) enforces the same physics in either CMS hemisphere.

\begin{figure*}[ttt]
\resizebox{1.\textwidth}{!}{ \includegraphics{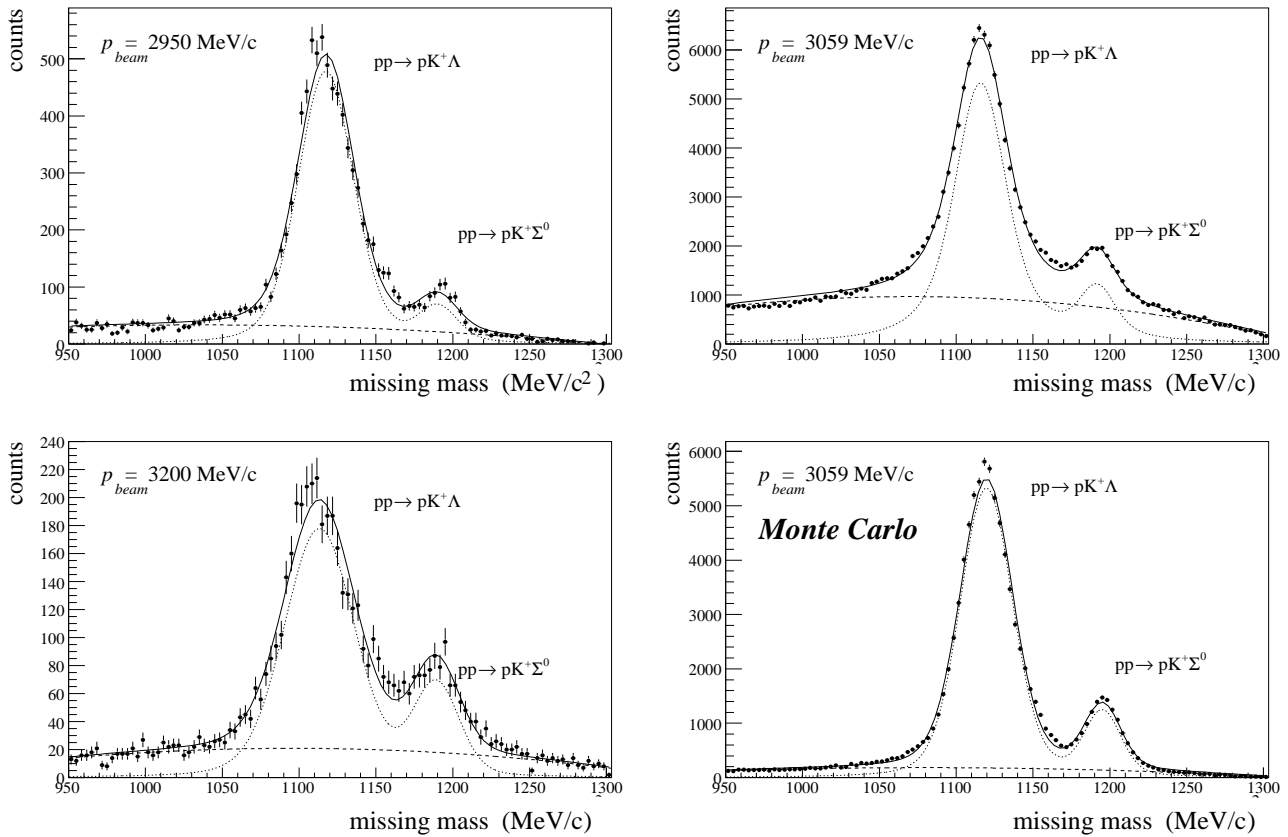} }
\caption{Missing mass spectra measured at the three beam momenta. 
Two peaks, corresponding to $\mathrm{pp \to pK^+\Lambda}$ and $\mathrm{pp\to pK^+\Sigma^0}$ are observed 
above a smooth background. The result of a fitting procedure for signal (dotted line), background (dashed line),
 and total 
spectrum (solid line) are shown in all cases. In the lower right frame the Monte Carlo result 
for 3059 MeV/c is depicted. 
The relative weight of both signals were accounted for by using the 
experimentally determined ratio of the total cross sections.}
\label{fig:viermalmm}  
\end{figure*}

\subsection{Acceptance correction and absolute normalization}
The Monte-Carlo package used \cite{lasvegasbeschreibungBrand,lasvegasbeschreibungZielinsky} 
models the detector and the physical processes to great detail. 
The event generator produces the particles of the exit channel either according to the three-body phase space,
or likewise, intermediate resonances can be chosen in order to model a two-step creation process 
($\mathrm{pp\to p}N^*, N^*\to\mathrm{K^+\Lambda}$, $\mathrm{K^+\Sigma^0}$).
The particles (and their daughters, granddaughters, ...) are then propagated through the detector. 
Branching ratios and lifetimes of all particles are incorporated according to the values given in \cite{pdg06}. 
Energy loss, small-angle scattering, nuclear reactions, and $\delta$-electrons are considered. 
From the energy deposit in the active detector components digitized QDC- and TDC-signals are generated. 
Noise and thresholds are modeled as known from the
measured detector response. 
Deviations from an homoge\-neous\-ly populated phase space can be introduced by a weight function on an event-by-event basis 
(this procedure is called {\it filtering} in the following). 
Finally, the Monte Carlo data are subjected to the very same routines as real data in order to determine the acceptance. 

The use of phase-space distributed data as Monte Carlo input is only justified if the reaction under study is
homogeneously distributed in phase space or if the acceptance coverage of the detector is homogeneous 
over the whole five-dimensional event-space of the three-body final state. 
Two deviations from three-body phase space are likely to occur in proton-proton experiments: 
anisotropic angular distributions causing particles to prefer different angular regions 
(of possibly different acceptance) and intermediate nucleon resonances
limiting the available phase space of all decay products caused by the mass and width of the resonance.

The influence of anisotropic angular distributions on the overall acceptance was investigated by 
filtering the Monte Carlo input in such a way as to match the experimental results.
For all modeled angular distributions a change of the overall acceptance of less than 10\% was found. 
The influence of nucleon resonances on the acceptance corrected data
was deduced by using a set of Monte Carlo data created with different $N^*$ resonances 
($m_{N^*}= 1400$, 1535, 1650, 1720, 1850, 1900 $\mathrm{MeV}$; 
$\Gamma_{N^*}=20$, 100, 150, 200, 300, 400 $\mathrm{MeV}$). 
Only minor changes of the acceptance corrected differential distributions were observed and the total cross section 
changes by less than $4\%$ if the width of the resonance was chosen to be larger than 100 MeV.
As, at present, there is no theoretical model available on which a proper Monte Carlo simulation can be based,
a homogeneously populated phase space modified to model measured angular distributions was used 
throughout the analysis. Details will be given when discussing the angular distributions.

The overall acceptance for the reactions 
under study is mainly governed by three 
obvious contributions, namely the $\mathrm{\Lambda}$ branching ratio to charged particles (64\%), the restriction to one CMS 
hemisphere (50\%), 
and the probability of the secondary vertex to be located within the 
``decay volume'' (30\%) indicated in the lower part of fig.~2.
The decay of kaons and pions in flight as well as a decay pion escaping detection contribute in an 
intertwined manner ($\approx25\%$). 
The overall acceptance is found to be $\approx 1.8$\% in case of $\mathrm{\Lambda}$ and $\approx 1.4$\% in case of 
$\mathrm{\Sigma^0}$ detection.
The relative uncertainty of the acceptance correction was determined by investigating the effect of all restrictions imposed 
during the data analysis and was found to be below 10\%. 

In the case of differential distributions the acceptance varies smoothly with the observable under consideration.
Here, an additional uncertainty comes into play due to the gradient of the acceptance function. 
This additional uncertainty $\Delta a_i$ is taken into account by choosing 
$\Delta a_i =(|a_i-a_{i-1}|+|a_i-a_{i+1}|)/4$, where $a_i$ is the acceptance in bin $i$. 
The square root of the quadratic sum of this acceptance uncertainty, the 
statistical error, and the uncertainty due to signal-background separation (see below) 
will be shown when presenting the data. 

The absolute normalization is determined via the ana\-lysis of elastic scattering, 
which was recorded simultaneously during the experiment.
Our results are normalized to high quality data on elastic cross sections provided 
by the EDDA collaboration \cite{edda00} and yielded the numerical
values for the time-integrated luminosity of $16.9\,\mathrm{nb^{-1}}$ (2950 MeV/c), $214\,\mathrm{nb^{-1}}$ (3059 MeV/c), 
and $6.4\,\mathrm{nb^{-1}}$ (3200 MeV/c).
The total uncertainty of this procedure (5\%) is in equal parts due to our analysis and the 
uncertainty of the literature data. 
For details see \cite{MSWdiss,WU07}.

\subsection{Determination of the signal yields}
Figure \ref{fig:viermalmm} shows the missing mass spectra measured at the three excess energies.
In addition the Monte Carlo result is shown for a beam momentum of 3059 MeV/c (lower-right)
The ratio of  
Monte Carlo $\mathrm{pK^+\Lambda}$ and $\mathrm{pK^+\Sigma^0}$ events are chosen according to 
the total cross sections measured for each excess energy.
Distinct signals for the $\mathrm{\Lambda}$ and the $\mathrm{\Sigma^0}$ hyperon can be seen above a smooth and 
structureless background. As usual for time-of-flight detectors,
the missing mass resolution (momentum resolution) 
is best for smaller velocities in the exit channel (smaller beam momenta in the entrance channel).
Comparing the spectra, the higher luminosity is reflected in the spectrum for 
$p_{beam}=3059\,\mathrm{MeV/c}$, a better beam quality for the measurements at 
$p_{beam}=2950$ and $3200\,\mathrm{MeV/c}$ manifests itself in a low background contribution.

The number of events in the missing-mass peaks are obtained by consecutively fitting 
first the  background and then the signals.
For the background parametrization qua\-dra\-tic or cubic polynomials are used, where only missing masses
below the $\mathrm{\Lambda}$- and above the $\Sigma^0$-peak are taken into account. 
Both types of parametrizations lead to the same final results within 2\%. 
Then, the background parameters are fixed and the signals are described by two Voigt functions 
(convolution of a Gauss- and a Lorentz-function).
Voigt functions are chosen since they model properly
the signal shape of a rather narrow peak accompanied by broader tails. 
A superposition of two Gaussians, however, yields the same results within uncertainty limits \cite{dietrich06}.

The overall systematic uncertainty due to signal and background separation is determined by varying 
the fit-region for the background-fit below and above the two peaks. 
In the case of $\mathrm{\Lambda}$ production this uncertainty is rather small (5\%), especially as
the contribution of the much smaller $\mathrm{\Sigma^0}$ signal is negligible.
The relative influence of the background on the systematic uncertainty is larger 
for the smaller $\mathrm{\Sigma^0}$-signal (15\%). 
Here, the signal-to-background ratio is close to unity and more sensitive to the choise of fitting regions. In addition,
the uncertainty of the contribution of the much larger $\mathrm{\Lambda}$ signal is not negligible, especially  
as the $\mathrm{\Sigma^0}$-signal happens to appear
near the maximum change of the slope of the $\mathrm{\Lambda}$-tail.

For the total cross sections statistical uncertainties which 
reflect the numerical uncertainties of the fitting procedure are quoted
while the systematic uncertainties are the quadratic sum of the uncertainties 
of luminosity determination (5\%), acceptance correction (10\%), and signal integration (5\% for $\mathrm{\Lambda}$ and 15\%
for $\mathrm{\Sigma^0}$).

Differential cross sections are determined in analogy to the total yield,
only that the amount of signal and background is determined individually from  missing mass spectra 
generated for each bin of the observable under study. 
The widths of the bins are chosen according to the detector resolution and statistical aspects.
In addition, the secondary $V$ is required to have in each arm four 
track points in the two hodoscopes. This requirement further reduces background and allows 
to separate a clean signal also in angular bins containing only a small number of counts.

\section{Results and discussions}
The interpretation of experimental data on proton-pro\-ton induced hyperon production 
is difficult for three reasons.
First off, the different reaction mechanisms (light/heavy non-strange as well 
as strange meson exchange, nucleon resonances, 
and FSI, ...) are likely to interfere, i.e.~the effects of the different contributions cannot be seen in a 
pure and isolated manner. Second, each particle in a three-body final state 
is always connected kinematically to the other two. This can lead to correlations between
two different observables and renders a true physical cause difficult to disentangle from 
its effect.
 A well known example for this to happen is the reflection of a resonance seen in a Dalitz plot. 
And finally, the interpretation of the experimental data by theory is often ambiguous as different conclusions
can be drawn from the same experimental data set. 
For these reasons the interpretation of the results presented will be using only  
general arguments based on kinematics and conservation laws.

In the following the total cross sections for both channels will be presented and discussed, 
where in particular the energy dependence ot the ratio $R_{\mathrm{\Lambda/\Sigma^0}}$
will be addressed. Then the differential distributions of the reaction 
$\mathrm{pp\to pK^+\Lambda}$ will be shown for all three excess energies 
in various reference frames. 
Using these results we will elaborate on the reaction mechanism as it manifests 
itself in these differential distributions. 
Finally, the first differential observables for 
$\mathrm{\Sigma^0}$ production in proton-proton collisions will be presented for the high statistics data set. 
Here the discussion will concentrate on a comparison of the two reaction channels and the reaction mechanisms involved. 
The numerical values of all one-dimensional cross sections are listed in the appendix.

\begin{table}
  \caption{Total cross sections for the reactions $\mathrm{pp\to pK^+\Lambda}$ and $\mathrm{pp\to pK^+\Sigma^0}$.
  The first uncertainty refers to statistical and the second to systematical ones.}
	\label{tab:totalcrosssec}
      \begin{tabular}{@{}llll}
      \hline\noalign{\smallskip}
      $\varepsilon\;(\mathrm{MeV})$   &  acc (\%) & counts & $\sigma_{\mathrm{tot}}\;(\mathrm{\mu b})$ \\
      \noalign{\smallskip}\hline\noalign{\smallskip}
	  $\mathrm{pp \to pK^{+}\Lambda}$ \\
      $204$  & 1.95 &  7228 &  \makebox[0.5cm][r]{ 21.8}  $\pm$ 0.3 $\pm$ 2.7 \\ 
      $239$  & 1.72 & 89684 &  \makebox[0.5cm][r]{ 24.4}  $\pm$ 0.1 $\pm$ 3.0 \\
      $284$  & 1.63 &  3322 &  \makebox[0.5cm][r]{ 32.0}  $\pm$ 0.9 $\pm$ 3.9 \\
      $\mathrm{pp \to 	pK^{+}\Sigma^0}$ \\
      $127$  & 1.28 &   676 &  \makebox[0.5cm][r]{  3.1}  $\pm$ 0.2 $\pm$ 0.6 \\
      $162$  & 1.51 & 12644 &  \makebox[0.5cm][r]{  3.9}  $\pm$ 0.1 $\pm$ 0.7 \\
      $207$  & 1.45 &   800 &  \makebox[0.5cm][r]{  8.6}  $\pm$ 0.5 $\pm$ 1.6 \\
      \noalign{\smallskip}\hline
    \end{tabular}
\end{table}

\subsection{Total cross sections}
\label{sigmatot}
\begin{figure}[hht]
\resizebox{0.5\textwidth}{!}{ \includegraphics{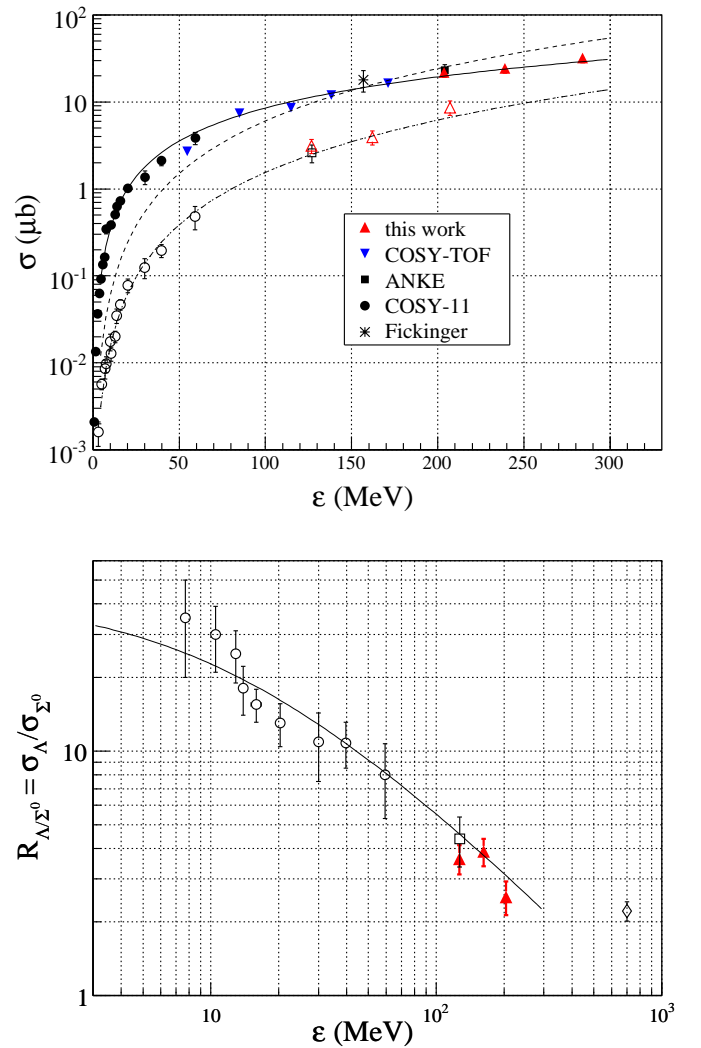} }
\caption{Upper part: the world data set on total cross sections of 
$\mathrm{pp\to pK^+\Lambda}$ (solid symbols) and $\mathrm{pp\to pK^+\Sigma^0}$ (open symbols) 
\cite{grzonka97,balewski97,balewski98,sewerin99,COSY11kowina04,toflambda06,valdau07,fickinger62}; dashed lines: phase space, solid line: phase space + FSI.  
  Lower part: energy dependence of the cross section ratio
 $R_{\mathrm{\Lambda/\Sigma^0}}$. The open diamond at $\epsilon = 700\,\mathrm{MeV}$ is an average value calculated 
 from the data given in
 \cite{baldini88}. The results of the present work are shown as triangles. The solid line is the ratio 
 of the parametrizations of the two excitation functions.}
\label{fig:worlddatatotalX}  
\end{figure}

The results for the total cross sections are listed in ta\-ble \ref{tab:totalcrosssec}. 
They are included in fig.~\ref{fig:worlddatatotalX} 
which shows in the upper part the world data of total cross sections 
for the reaction $\mathrm{pp\to pK^+\Lambda}$ and $\mathrm{pp\to pK^+\Sigma^0}$  ($\epsilon<300$ MeV). 
Our new data extend the near threshold measurements by roughly 80 MeV excess energy, i.e.~directly 
into the region where the energy dependence of the 
ratio of the cross sections $R_{\mathrm{\Lambda/\Sigma^0}}$ is not known well. 
Up to $\epsilon\approx 170\,\mathrm{MeV}$ the excitation function 
$\sigma_{\mathrm{\Sigma^0}}=\sigma_{\mathrm{\Sigma^0}}(\epsilon)$ 
is very well described within the experimental resolution by a pure phase space dependence 
given by $\sigma_{\mathrm{pK^{+}\Sigma^0}}= K \cdot \epsilon^2$ 
($K=1.545\times 10^{-4} \mu b/MeV^2$, dash-dotted line in fig.~\ref{fig:worlddatatotalX}).

The excitation function of $\mathrm{pp\to pK^+\Lambda}$ clearly cannot be parameterized by pure phase space 
(dashed line in fig.~\ref{fig:worlddatatotalX}). A parametrization proposed by 
F\"aldt and Wilkin \cite{faeldtwilkin97} is more appropriate
\begin{equation}
\sigma = C \cdot \frac{\epsilon^2}{(1+ \sqrt{1+\epsilon/\alpha})^2},
\label{func:faeldwilkin}
\end{equation}
and describes the energy dependence of the cross section well up to 300 MeV (solid line in 
fig.~\ref{fig:worlddatatotalX}; the parameters $C=0.02574\mu b/MeV^2$ 
and $\alpha=5.203 MeV$ can be related to the $\mathrm{p\Lambda}$-FSI).

In the lower part of fig.~\ref{fig:worlddatatotalX} the ratio of both parametrizations is shown as a solid 
line together with the 
world data on cross section ratios \cite{balewski98,COSY11kowina04,baldini88}. 
At the excess energies of 127 MeV a data point is added which is determined from the 
measured $\sigma_{\mathrm{\Sigma^0}}$ cross sections given in \cite{valdau07} and 
the $\sigma_{\mathrm{\Lambda}}$ parametrization (\ref{func:faeldwilkin}). 
At the same energy and at 162 MeV  
our results (shown as triangles) are also included, where the $\sigma_{\mathrm{\Lambda}}$ 
parametrization (\ref{func:faeldwilkin}) and our new  cross sections 
for $\sigma_{\mathrm{\Sigma^0}}$ are used.

The two experimental cross sections of $\mathrm{pp\to pK^+\Lambda}$ at $\epsilon=204\,\mathrm{MeV}$ and 
$\mathrm{pp\to pK^+\Sigma^0}$ at $\epsilon=207\,\mathrm{MeV}$ are used  to directly  calculate the ratio
of $R_{\mathrm{\Lambda/\Sigma^0}}=2.5\pm 0.4$, 
as the phase space volume differs by less than 3\%.
In this ratio systematic uncertainties of the two measurements partly cancel (see introduction). 
The new experimental value confirms the general trend 
towards the high energy limit of 2.2, which has been determined experimentally for 
$\epsilon>700\,\mathrm{MeV}$ in the 1960ies and 1970ies. At an excess energy of 
204 MeV $R_{\mathrm{\Lambda/\Sigma^0}}$ is found to be more than one standard deviation below the 
value of three which follows from isospin considerations, and thus, this reasoning is 
unlikely to be the proper approach 
to explain the $\mathrm{\Lambda}$ to $\mathrm{\Sigma^0}$ cross section ratio.

Obviously, the energy region of enhanced $\mathrm{\Lambda}$ over $\mathrm{\Sigma^0}$ production 
ends at excess energies of slightly above $\epsilon\approx 200\,\mathrm{MeV}$. 
This surely is a surprise and will have implications
on theory as for example the calculations of \cite{sibi06} predict values of  $R_{\mathrm{\Lambda/\Sigma^0}}\approx5$ 
at $\epsilon=200\,\mathrm{MeV}$ and must be reconciled with the present finding. 
The same authors, however, point out that the total cross sections for hyperon production  
are not sufficient to tightly constrain the parameters of any model.
Therefore, differential distributions will be presented next.

\subsection{Differential distributions: $\mathrm{pp}\to \mathrm{pK^{+}\Lambda}$}
Differential data for the reaction $\mathrm{pp\to pK^+\Lambda}$ are presented in figs.~\ref{fig:dalitz}
to \ref{fig:finalbild}.
In the one-dimensional differential distributions, data points are included as long as 
the uncertainty (quadratic sum of statistical error, the uncertainty of signal integration and 
acceptance correction) in a specific bin is below 80\% of its cross section value. 
In order to base the discussion on a quantitative footing, all one-dimensional distributions have 
been subjected to a least square fitting with Legendre polynomials
${\mathrm d}\sigma/{\mathrm d}\Omega=\sum _{l=0}^{l_{max}} a_l\cdot P_l,\quad{l=0,1,2,4}$. 
The coefficients are listed in the respective tables \ref{tab:fitparasL1} to \ref{tab:fitparasL3} and 
will be used only to judge anisotropies ($P_2$ and $P_4$) and asymmetries ($P_1$, representative for 
all $P_{odd}$). The energy dependence of the
observable under study can also be inferred and is found to be, in general, rather weak.
It should be noted in passing that the total cross section given by the integral of each differential distribution 
($\sigma_{tot} = \int \frac{d\sigma}{d\Omega}d\Omega = 4\pi\cdot a_0$) in all cases is compatible within uncertainty
with the values listed in table \ref{tab:totalcrosssec}.
The results obtained from (filtered) Monte Carlo data and the detector acceptance will be shown where appropriate.

\subsubsection{Dalitz plots}
\begin{figure}[ttt]
\resizebox{0.5\textwidth}{!}{ \includegraphics{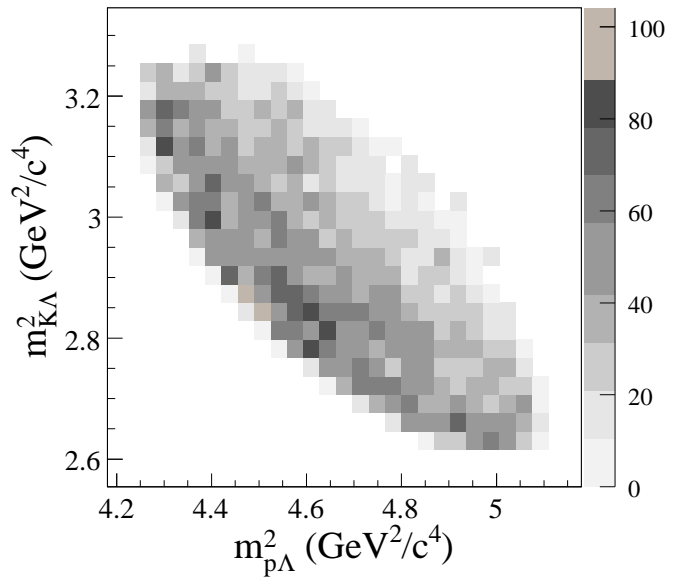} }
\caption{Dalitz plot of the $\mathrm{pK^+\Lambda}$ final state measured at $\epsilon = 204\,\mathrm{MeV}$.
The data is acceptance corrected, however the background is not subtracted (see text). 
The scale of the relative bin occupancy is shown at the right side.
}
\label{fig:dalitz}  
\end{figure}
The acceptance corrected Dalitz plot ($m_{\mathrm{K\Lambda}} \mathrm{vs.} ~m_{\mathrm{p\Lambda}}$) 
of the $\mathrm{pK^+\Lambda}$ final 
state is shown in fig.~\ref{fig:dalitz} for an excess energy of 204 MeV ($p_{beam} = 2950\,\mathrm{MeV/c}$). 
Data is shown for a missing mass region of $\pm 50\,\mathrm{MeV/c^2}$ around the $\mathrm{\Lambda}$ mass,
hence reducing the contribution of background and $\mathrm{\Sigma^0}$ hyperons.
The plot is not corrected for background, 
however, sideband cuts below the $\mathrm{\Lambda}$-mass have been checked 
and show the background not to be responsible for the structures seen.

The relative bin occupancy of the Dalitz plot shown in fig.~\ref{fig:dalitz} resembles strongly the 
one presented in \cite{toflambda06}, which was determined for a smaller
excess energy of $\epsilon=171\,\mathrm{MeV}$  ($p_{beam} = 2850\,\mathrm{MeV}$).
In both cases the kinematically allowed region is covered,
an enhancement of the data is apparent along the lower $\mathrm{p\Lambda}$-mass boundary, and
no prominent resonant band  is observed along the $\mathrm{m^2_{p\Lambda}}$ axis.
This signature is explained 
in \cite{toflambda06} by means of a quantitative Dalitz plot analysis. 
It was found that the nucleon resonances N(1650), N(1710), and N(1720) in conjuction with 
a sizable p$\mathrm{\Lambda}$ FSI play a decisive role, however, 
both are strongly interrelated by interference effects. 
A detailed analysis of a series of Dalitz plots measured at excess energies of 204, 284, and 316 MeV
is the subject of a forthcoming publication of the {\it COSY-TOF} collaboration \cite{erlangenzukunft10}.

The present paper, hence, does not aim at an analysis of the Dalitz plot.
We rather focuses on presenting and 
discussing sets of one-dimensional differential distributions 
which have not been published so far.
Our results substantially complement as well as support the earlier Dalitz plot analsis 
of \cite{toflambda06}.

\begin{figure}[ttt]
\resizebox{0.5\textwidth}{!}{ \includegraphics{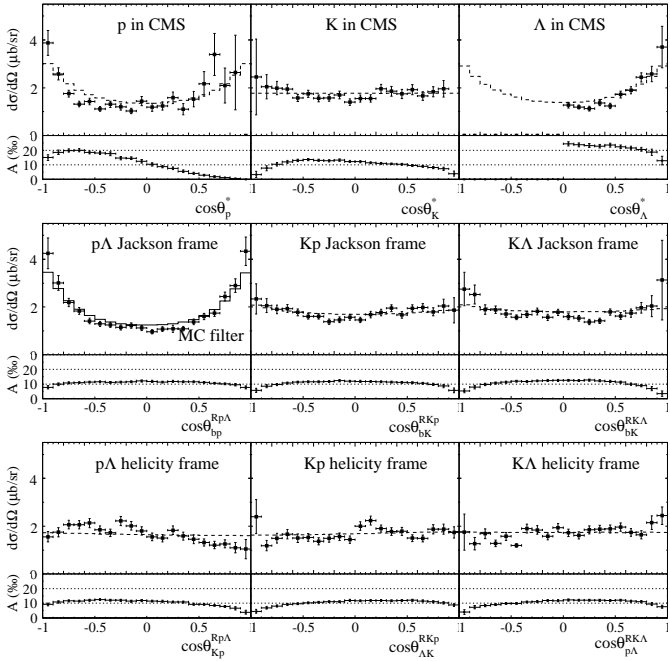} }
\caption{Angular distributions of the particles in the overall CMS, Jackson and
helicity frames (top-down) for the reaction $\mathrm{pp\to pK^{+}\Lambda}$ 
measured at an excess energy of $\epsilon=204\,\mathrm{MeV}$ ($p_{beam}=2950\,\mathrm{MeV/c}$).
Error bars for each data point are  the square root of the quadratic sum of the statistical, acceptance, and 
signal-to-background-separation uncertainty.
The solid histogram in the $\mathrm{p\Lambda}$ Jackson frame represents the Legendre polynomial 
of table \ref{tab:fitparasL1} which is used as MC filter. Its effect on all other angular distributions is shown
by the dashed histograms. Below each angular distributions the differential acceptance is shown. 
}
\label{fig:2950lambda}  
\end{figure}

\begin{figure}[ttt]
\resizebox{0.5\textwidth}{!}{ \includegraphics{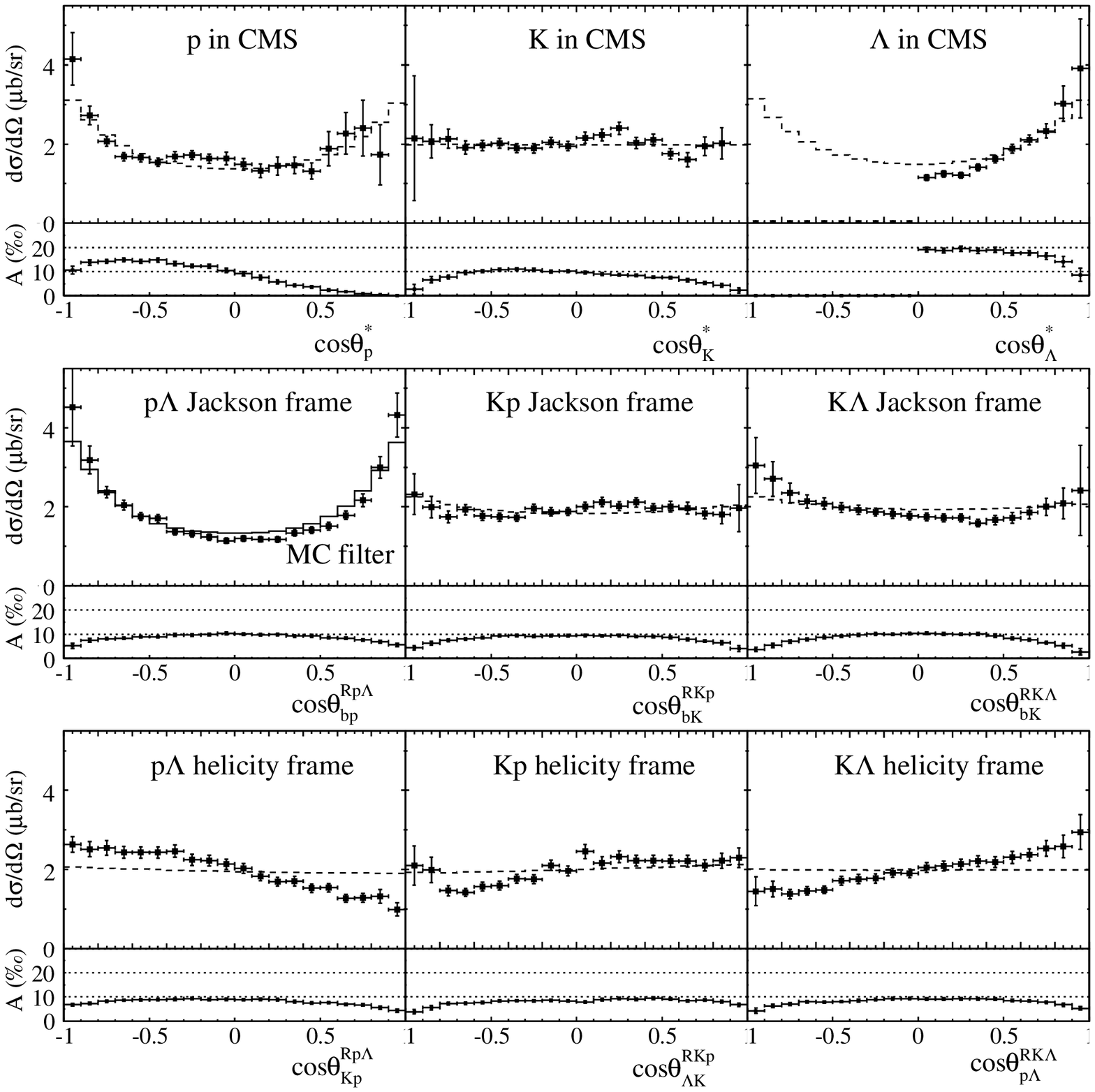} }
\caption{Angular distributions of the particles in the overall CMS, the Jackson and
helicity frames (top-down) for the reaction $\mathrm{pp\to pK^{+}\Lambda}$ 
measured at an excess energy of $\epsilon=239\,\mathrm{MeV}$ ($p_{beam}=3059\,\mathrm{MeV/c}$).
Error bars for each data point are the square root of the quadratic sum of the statistical, acceptance, and 
signal-to-background-separation uncertainty.
The solid histogram in the $\mathrm{p\Lambda}$ Jackson frame represents the Legendre polynomial 
of table \ref{tab:fitparasL2} which is used as MC filter. Its effect on all other angular distributions is shown
by the dashed histograms. Below each angular distributions the differential acceptance is shown. 
}
\label{fig:3059lambda}  
\end{figure}

\begin{figure}[thh]
\resizebox{0.5\textwidth}{!}{ \includegraphics{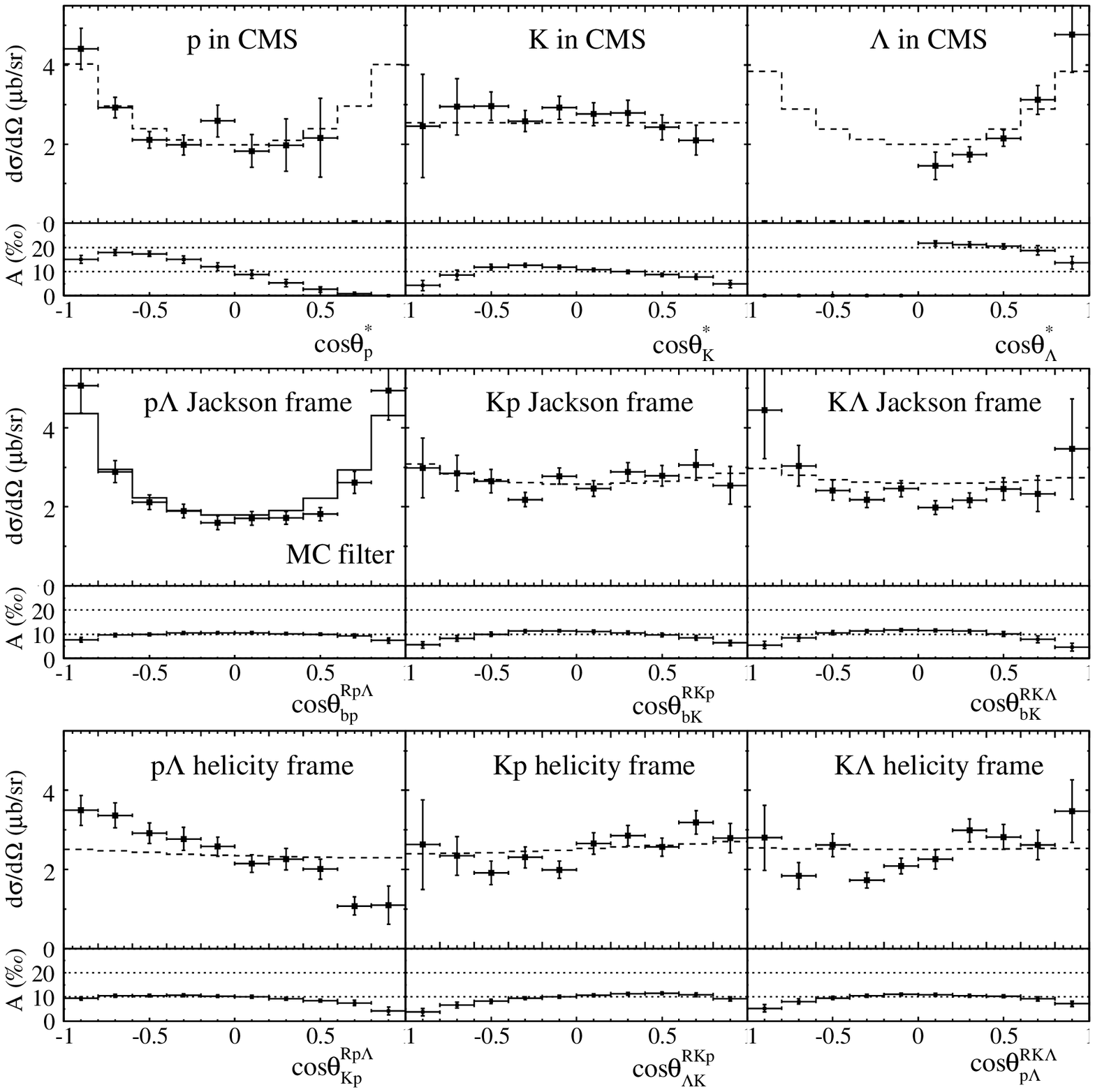} }
\caption{Angular distributions of the particles in the overall CMS, the Jackson and
helicity frames (top-down) for the reaction $\mathrm{pp\to pK^{+}\Lambda}$ 
measured at an excess energy of $\epsilon=284\,\mathrm{MeV}$ ($p_{beam}=3200\,\mathrm{MeV/c}$).
Error bars for each data point are the square root of the quadratic sum of the statistical, acceptance, and 
signal-to-background-separation uncertainty.
The solid histogram in the $\mathrm{p\Lambda}$ Jackson frame represents the Legendre polynomial 
of table \ref{tab:fitparasL3} which is used as MC filter. Its effect on all other angular distributions is shown
by the dashed histograms. Below each angular distributions the differential acceptance is shown. 
}
\label{fig:3200lambda}  
\end{figure}

\begin{table}[ttt]
\caption{Coefficients of Legendre polynomials (in units of $\mu $b/sr) determined by least square 
fitting to angular distributions 
of the reaction $\mathrm{pp\to pK^+\Lambda}$ at
$\epsilon=204\,\mathrm{MeV}$ ($p_{beam}=2950\,\mathrm{MeV/c}$).} 
\label{tab:fitparasL1}
     	\begin{tabular}{@{}lrrrr}
  		\hline\noalign{\smallskip}
 cos & $a_0\;\;\;\;\;$ & $a_1\;\;\;\;\;$ & $a_2\;\;\;\;\;$  & $a_4\;\;\;\;\;$  \\   		\noalign{\smallskip}\hline\noalign{\smallskip}
$\theta^*_{p}$                  &   $ 1.87 \pm .12 $ &   $ 0.40 \pm .19 $ &	$ 1.85 \pm .34 $ &   $ 0.78 \pm .26 $    \\ \noalign{\smallskip}
$\theta^*_{K}$                  &   $ 1.76 \pm .06 $ &   $ 0.04 \pm .10 $ &	$ 0.37 \pm .16 $ &   ---$\;\;\;\;\;\,$   \\ \noalign{\smallskip}
$\theta^*_{\Lambda}$            &   $ 1.78 \pm .08 $ &   ---$\;\;\;\;\;\,$ &  $ 1.36 \pm .18 $ &   ---$\;\;\;\;\;\,$   \\ \noalign{\smallskip}\hline\noalign{\smallskip}
$\theta^{Rp\Lambda}_{bp}$       &   $ 1.77 \pm .06 $ &   $ -0.00 \pm .10 $ &   $ 1.59 \pm .14 $ &   ---$\;\;\;\;\;\,$   \\ \noalign{\smallskip}
$\theta^{RKp}_{bK}$             &   $ 1.79 \pm .04 $ &   $ 0.10 \pm .08 $ &	$ 0.50 \pm .12 $ &   ---$\;\;\;\;\;\,$   \\ \noalign{\smallskip}
$\theta^{RK\Lambda}_{bK}$       &   $ 1.76 \pm .05 $ &  $ -0.11 \pm .08 $ &	$ 0.45 \pm .13 $ &   ---$\;\;\;\;\;\,$   \\ \noalign{\smallskip}\hline\noalign{\smallskip}
$\theta^{Rp\Lambda}_{Kp}$       &   $ 1.64 \pm .04 $ &  $ -0.49 \pm .08 $ &  $ -0.32 \pm .11 $ &   ---$\;\;\;\;\;\,$   \\ \noalign{\smallskip}
$\theta^{RKp}_{\Lambda K}$      &   $ 1.61 \pm .04 $ &   $ 0.22 \pm .08 $ &  $ -0.14 \pm .11 $ &   ---$\;\;\;\;\;\,$   \\ \noalign{\smallskip}
$\theta^{RK\Lambda}_{p\Lambda}$ &   $ 1.62 \pm .04 $ &   $ 0.46 \pm .08 $ &  $ -0.16 \pm .13 $ &   ---$\;\;\;\;\;\,$   \\ \noalign{\smallskip} 
  	\noalign{\smallskip}\hline
	\end{tabular}
\end{table}

\begin{table}[ttt]
\caption{Coefficients of Legendre polynomials (in units of $\mu $b/sr) determined by least square fitting to angular distributions 
of the reaction $\mathrm{pp\to pK^+\Lambda}$ at
$\epsilon=239\,\mathrm{MeV}$ ($p_{beam}=3059\,\mathrm{MeV/c}$).} 
\label{tab:fitparasL2}
     	\begin{tabular}{@{}lrrrr}
  		\hline\noalign{\smallskip}
 cos & $a_0\;\;\;\;\;$ & $a_1\;\;\;\;\;$ & $a_2\;\;\;\;\;$  & $a_4\;\;\;\;\;$  \\   		\noalign{\smallskip}\hline\noalign{\smallskip}
$\theta^*_{p}$ 					&  $ 1.94 \pm 0.12 $ &   $ -0.16 \pm .15 $ &  $ 1.31  \pm .35 $ &	$ 0.90 \pm .26 $	 \\ \noalign{\smallskip}						  
$\theta^*_{K}$ 					&  $ 1.94 \pm 0.05 $ &   $ -0.01 \pm .09 $ &  $ -0.25 \pm .14 $ &	---$\;\;\;\;\;\,$	\\ \noalign{\smallskip} 						  
$\theta^*_{\Lambda}$ 			&  $ 1.92 \pm 0.08 $ &   ---$\;\;\;\;\;\,$ &  $ 1.54  \pm .19 $ &  ---$\;\;\;\;\;\,$   \\ \noalign{\smallskip}\hline\noalign{\smallskip}  
$\theta^{Rp\Lambda}_{bp}$ 		&  $ 1.86 \pm 0.05 $ &   $ -0.17 \pm .09 $ &  $ 1.47  \pm .12 $ &	---$\;\;\;\;\;\,$	\\ \noalign{\smallskip} 						  
$\theta^{RKp}_{bK}$ 			&  $ 1.89 \pm 0.04 $ &   $ 0.14 \pm .07 $ &   $ -0.11 \pm .10 $ &	---$\;\;\;\;\;\,$	\\ \noalign{\smallskip} 						  
$\theta^{RK\Lambda}_{bK}$ 		&  $ 1.96 \pm 0.05 $ &  $ -0.33 \pm .08 $ &   $ 0.46  \pm .13 $ &	---$\;\;\;\;\;\,$	\\ \noalign{\smallskip}\hline\noalign{\smallskip} 
$\theta^{Rp\Lambda}_{Kp}$ 		&  $ 1.95 \pm 0.03 $ &  $ -0.87 \pm .06 $ &   $ -0.16 \pm .08 $ &	---$\;\;\;\;\;\,$	\\ \noalign{\smallskip} 						  
$\theta^{RKp}_{\Lambda K}$ 		&  $ 1.91 \pm 0.03 $ &   $ 0.56 \pm .06 $ &   $ -0.28 \pm .10 $ &	---$\;\;\;\;\;\,$	\\ \noalign{\smallskip} 						  
$\theta^{RK\Lambda}_{p\Lambda}$ &  $ 1.96 \pm 0.04 $ &   $ 0.70 \pm .07 $ &   $ -0.01 \pm .10 $ &	---$\;\;\;\;\;\,$	\\ \noalign{\smallskip} 						  
  	\noalign{\smallskip}\hline
	\end{tabular}
\end{table}

\begin{table}[ttt]
\caption{Coefficients of Legendre polynomials (in units of $\mu $b/sr) determined by least square fitting to angular distributions 
of the reaction $\mathrm{pp\to pK^+\Lambda}$ at
$\epsilon=284\,\mathrm{MeV}$ ($p_{beam}=3200\,\mathrm{MeV/c}$).} 
\label{tab:fitparasL3}
		\begin{tabular}{@{}lrrrr}
		\hline\noalign{\smallskip}
 cos & $a_0\;\;\;\;\;$ & $a_1\;\;\;\;\;$ & $a_2\;\;\;\;\;$  & $a_4\;\;\;\;\;$  \\			\noalign{\smallskip}\hline\noalign{\smallskip}
$\theta^*_{p}$  				&   $ 2.74 \pm .52 $ &    $ -0.19 \pm .76 $ &  $ 2.29  \pm 1.45 $ &   $ 1.54 \pm 1.18 $    \\ \noalign{\smallskip}
$\theta^*_{K}$  				&   $ 2.55 \pm .18 $ &    $ -0.35 \pm .31 $ &  $ -0.49 \pm 0.50 $ &    ---$\;\;\;\;\;\,$   \\ \noalign{\smallskip}
$\theta^*_{\Lambda}$			&   $ 2.61 \pm .28 $ & ---$\;\;\;\;\;\,    $ &  $ 2.49  \pm 0.74 $ &   ---$\;\;\;\;\;\,$   \\ \noalign{\smallskip}\hline\noalign{\smallskip}
$\theta^{Rp\Lambda}_{bp}$		&   $ 2.60 \pm .19 $ &    $ -0.19 \pm .33 $ &  $ 2.16  \pm 0.43 $ &   ---$\;\;\;\;\;\,$   \\ \noalign{\smallskip}
$\theta^{RKp}_{bK}$ 			&   $ 2.67 \pm .12 $ &    $ 0.25  \pm .23 $ &	$ 0.25  \pm 0.33 $ &   ---$\;\;\;\;\;\,$   \\ \noalign{\smallskip}
$\theta^{RK\Lambda}_{bK}$		&   $ 2.59 \pm .18 $ &    $ -0.28 \pm .28 $ &	$ 0.97  \pm 0.44 $ &   ---$\;\;\;\;\;\,$   \\ \noalign{\smallskip}\hline\noalign{\smallskip}
$\theta^{Rp\Lambda}_{Kp}$		&   $ 2.34 \pm .10 $ &    $ -1.37 \pm .20 $ &  $ -0.23 \pm 0.24 $ &   ---$\;\;\;\;\;\,$   \\ \noalign{\smallskip}
$\theta^{RKp}_{\Lambda K}$  	&   $ 2.44 \pm .12 $ &    $ 0.65  \pm .24 $ &	$ 0.17  \pm 0.35 $ &   ---$\;\;\;\;\;\,$   \\ \noalign{\smallskip}
$\theta^{RK\Lambda}_{p\Lambda}$ &   $ 2.48 \pm .15 $ &    $ 0.73  \pm .25 $ &	$ 0.65  \pm 0.45 $ &   ---$\;\;\;\;\;\,$   \\ \noalign{\smallskip} 
	\noalign{\smallskip}\hline
	\end{tabular}
\end{table}

\subsubsection{Angular distributions in the overall CMS}
The angular distributions of the three ejectiles in the overall CMS are shown in the upper row
of figs.~\ref{fig:2950lambda} to~\ref{fig:3200lambda}. The dashed histograms  shown in the same pictures 
correspond to filtered Monte Carlo data and will be explained in detail in the following 
section. The change of acceptance is smooth as shown below each distribution. 
The steep decline of acceptance towards
$\mathrm{cos}\theta=1$ in the case of the kaon and proton distributions is 
caused by the requirement of evaluating only hyperons emitted into the forward CMS hemisphere.
Nevertheless, the angular distributions of proton and kaon are symmetric 
with respect to $\mathrm{cos}\theta=0$ as the coefficients 
$a_1$ in table \ref{tab:fitparasL1} to \ref{tab:fitparasL3} are compatible with zero. 
This symmetry in the overall CMS is mandatory in the case of identical particles 
in the entrance channel. 

The angular distributions of protons and $\mathrm{\Lambda}$ hyperons show a pronounced 
anisotropy. The proton distributions even require the inclusion of $P_4$. 
These anisotropies reflect relative angular 
momentum, $L$, in both the $\mathrm{p-(K^+\Lambda)}$ and $\mathrm{\Lambda-(K^+p)}$ system. 
From an inspection of the Legendre polynomial coefficients $a_l$ in tables~\ref{tab:fitparasL1} to \ref{tab:fitparasL3} 
one deduces ($L \leq 2$) 
for the former and ($L \leq 1$) for the latter. In contrast, the angular distributions of the kaons 
are essentially isotropic with $a_2$ coefficients deviating slightly from zero only
for the two lower energies. This indicates relative angular momentum in the $\mathrm{K^+-(p\Lambda)}$ 
system, if at all present, to become of even minor importance with increasing excess energy. 
 
The three CMS angular distributions are compatible with each other simply due to kinematics.
If the $\mathrm{K^+\Lambda}$ system is assumed, as working hypothesis, to form an $N^*$-resonance the angular 
distribution of $N^*$ is the mirror image of that of the associated proton. The $N^*$-resonance decays 
in its own rest frame back to back into  hyperon and kaon. The available energy in this two-body decay depends on the 
mass of the resonance and may vary between zero and $\epsilon$. However, due to
the phase space volume, it will mainly be at intermediate excess energies, i.e. in the order of 100 MeV. 
Kinematics then constrains the 
(heavier) hyperon to mainly preserve the direction of flight of the $N^*$-resonance, while the (lighter) kaon
can be emitted isotropically. This kinematic situation holds also true if the $N^*$
resonance has a very large width; then the kinematics becomes similar to that of particles 
distributed homogeneously according to three-body phase space.

Thus, the angular distributions in the overall CMS are not well suited to directly draw conclusions
on resonant or non-resonant production, as the former always displays a convolution of a two step 
process ($\mathrm{pp\to p}N^*$, $N^*\to \mathrm{K^+\Lambda}$).  
It will be shown in the following that Jackson and helicity frames are the much more natural choice of Lorentzian frames
in order to study the reaction properties due to intermediate resonances.

\subsubsection{Angular distributions in Jackson frames}
The motivation for an analysis within Jackson frames\footnote{\label{GJerklaerung} For reactions of type $ab\to \it 123$ the Jackson frame is defined as the
Lorentzian frame in which the center of mass of the particles {\it (2,3)} is at rest ($\vec{p}_3=-\vec{p}_2$). In this frame
the Jackson angle is defined as the angle between the beam direction and that of particle {\it 3}, 
i.e.~$\sphericalangle(\vec{p_b},\vec{p_3})$. This frame connects exit and entrance channel and carries information not accessible 
by means of a Dalitz plot analysis.
In a three-body final state three two-body subsystems can be used to define a Jackson frame ({\it {R23, R31, R12}}; 
$R$ indica\-ting $R$est frame). The nomenclature of \cite{Byckling} is adopted: For $\theta^{\mathrm{R23}}_{\mathrm{b3}}$ 
the superscripts denotes the 
rest frame under consideration, the subscripts indicate the angle of particle {\it 3} with respect to the beam $b$.}, 
introduced by Gottfried and Jackson as early as 1964 \cite{GJ}, 
arises by considering the  $\mathrm{\pi p\to K}Y$ vertex in 
fig.~\ref{fig:ractionmechnism}b, c to represent an isolated ``$2 \to 2$'' reaction 
(the following argumentation is similar 
for the  $\mathrm{Kp\to Kp}$ vertex in fig.~\ref{fig:ractionmechnism}a).
In this picture, the inverse reaction ($\mathrm{K}Y\mathrm{\to  p\pi}$) must have the same properties 
due to time reversal invariance.
If one now imagines colliding beams of kaons and hyperons with $\vec{p}_{Y} = -\vec{p}_{\mathrm{K}}$, 
which is by definition the $\mathrm{K}Y$ Jackson frame, it is self-evident that the distribution of angles 
$\theta^{\mathrm{RK}Y}_{\mathrm{bK}}$ 
of the (in this case emerging) proton with respect to the 
(in this case beam-axis defining) kaon directly gives information on the relative angular momenta involved.
This argumentation holds with or without an intermediate nucleon resonance.

Considering only this vertex four different particles are involved and, hence, 
the angular distributions are not at all restricted to 
show any symmetry with respect to $\mathrm{cos}\theta^{\mathrm{RK\Lambda}}_{\mathrm{bK}}=0$. 
In fact, $\mathrm{\pi^- p \to K^0\Lambda}$ scattering shows a strong anisotropy which can be traced back 
to the interference of 
resonances with opposite parity \cite{knasel75,baker78}. 
Of course, the concept of an analysis within a Jackson frame is based on the presumption 
that there is no difference between a free and a virtual pion, 
and the form factor entering the $\mathrm{\pi p\to K}Y$ vertex is the same for both the two-body and the 
three-body reaction $\mathrm{pp\to pK}Y$ with no influence of the additional proton in the latter case.

A peculiarity arises in our case from the fact that the entrance channel consists of 
identical particles which results in symmetric angular distributions in the overall CMS in which
 beam and target are collinear. 
 However, this collinearity is destroyed 
when the system is boosted into a Jackson frame (the angle between the two protons is $\approx160^\circ$  on average). 
As beam and target particle are indistinguishable, the reference axis for the Jackson angle 
can be the direction of either proton.
Therefore, the quantum mechanical identity of beam and target proton 
enforces the {\it same} angular distributions in the Jackson frame, when measured with respect to
either proton. However, as a Jackson frame is some other Lorentzian frame than the CM-system, the distributions
are not required to show {\it any} symmetry. 

The angular distributions in the Jackson frames 
are shown in the middle rows of figs.~\ref{fig:2950lambda} to \ref{fig:3200lambda}.
As the two protons of the initial state cannot be distinguished, 
both Jackson angles with respect to beam and target have been taken into account for each event 
(maintaining for simplicity the subscript $b$ in $\theta^{\mathrm{R23}}_{\mathrm{b3}}$). 

A pronounced anisotropy is observed in the $\mathrm{p\Lambda}$ Jackson frame
which is due to a relative angular momentum of $L=1$ in the $\mathrm{p\Lambda}$ system
(see table \ref{tab:fitparasL1} to \ref{tab:fitparasL3}).
This anisotropy as well as those of the angular distributions of protons and $\Lambda$-hyperons in the overall CMS
suggest their connection through kinematics.
In order to investigate
this conjecture all distributions measured in the CMS and the Jackson frames 
were used as weight functions for Monte Carlo simulations.
It was found that solely the filter on the distribution in the $\mathrm{p\Lambda}$ Jackson frame (solid histogram in
figs.~\ref{fig:2950lambda} to~\ref{fig:3200lambda}) results in a consistent and
satisfactory  description of all CMS distributions, illustrated by the dashed histograms.
It is concluded that the resonance reaction
$\mathrm{pp \to p}N^*$ with angular momentum in this ``exit'' channel is of importance. 
The $\mathrm{p}N^*$ rest frame (which coincides with the CMS) is well represented by the $\mathrm{p\Lambda}$ Jackson
frame, due to the large mass difference of kaon and hyperon.
Hence the $\mathrm{p\Lambda}$ Jackson frame is the natural reference frame to study the dynamics of the intermediate
$\mathrm{p}N^*$ system.
It should be stressed that this information on angular momentum ($L\leq2$) in the p-$N^*$ system is 
accessible neither by inspecting the CMS distributions nor the Dalitz plot.
It is explicitly the choice of the respective Jackson frame which enables one
to identify this aspect of the reaction dynamics.

The angular distributions in the K$^+$p-Jackson frame are expected to be basically isotropic  
if the reaction procedes in two steps via an intermediate $N^*$-resonance as the final state 
protons and kaons do not originate from the same vertex (cf. fig.~\ref{fig:ractionmechnism}b,c) 
and are correlated only through kinematics of the final state. If, however, kaon exchange as 
indicated in fig.~\ref{fig:ractionmechnism}a plays an important role the distribution in the 
K$^+$p-Jackson frame could reflect properties of that process which may even involve a yet 
unknown intermediate pentaquark-resonance. The experimental angular distributions are 
symmetric within the experimental uncertainties (cf. $a_1$ coefficients in 
tables~\ref{tab:fitparasL1} to~\ref{tab:fitparasL3}). 
There is a clear tendency for the coefficient $a_2$ to decrease with increasing 
excess energy pointing at an angular momentum $L$ of at most one unit to be 
present in the K$^+$p interaction at smaller excess energy while $L$ tends 
towards zero at higher excess energies. The Monte Carlo angular distributions 
resulting from the filter applied in the $\mathrm{p\Lambda}$ Jackson frame 
reproduce the data quite well. This is taken as evidence for a kinematical 
correlation rather than a strong indication for kaon exchange, which, however, 
cannot be excluded to contribute at some level. In fact, Balestra {\it et al.} \cite{balestra99} 
concluded from negative spin transfer coefficients $\mathrm{D_{NN}}$,
measured in exclusive $\mathrm{\Lambda}$ production from $\mathrm{p}\overrightarrow{\mathrm{p}}$  collisions at 3.67 GeV/c, kaon 
exchange to contribute to the reaction 
mechanism in conjuction with pion exchange and $\mathrm{\Lambda p}$ final state interaction. 

The experimental angular distributions in the $\mathrm{K^+\Lambda}$ Jackson frame 
tend to be both anisotropic and asymmetric. 
Both effects are only poorly reproduced by filtered phase-space Monte Carlo data. 
It is thus tempting to assume these angular distributions to be caused by the excitation 
of $N^*$-resonances decaying 
into the $\mathrm{K^+\Lambda}$ channel.
All these nucleon resonances 
have large widths and may contribute through their broad tails to the reaction, even if their central mass is 
outside the mass region between the threshold (1609 MeV/c$^2$) and 1893 MeV/c$^2$ (for the highest excess energy).
An inspection of \cite{pdg06} reveals a long list of possibly participating $N^*$ resonances, 
classified as either $\mathrm{S_{11}}$, 
$\mathrm{P_{11}}$, $\mathrm{P_{13}}$, $\mathrm{D_{13}}$, $\mathrm{D_{15}}$, or $\mathrm{F_{15}}$. 
A $\Delta^*$-resonance cannot contribute due to isospin conservation.
The coefficients of table~\ref{tab:fitparasL1} to \ref{tab:fitparasL3}
show $a_1$ to have a tendency to be non-zero while $a_2$ is non-zero. 
The fact that the inclusion of Legendre 
polynomials up to $\mathrm{P_2}$ suffices for a good description shows that only angular momenta of 
$L\leq 1$ are participating in this ``2$\to$2 reaction'' or, in other words, if $N^*$-resonances are 
involved their decay angular momentum must be $L\leq 1$. 
This constrains the resonances possibly involved to $\mathrm{S_{11}}$, $\mathrm{P_{11}}$, and $\mathrm{P_{13}}$, 
where $L=0$ belongs to the former, $L=1$ to the latter two. Contributions of 
$D_{13}$, $D_{15}$, or $F_{15}$ resonances, which would require $L>1$, are therefore not supported by the data.

The angular distribution of a true two-body resonance reaction is asymmetric ($a_1 \neq 0$) only if resonances 
with both parities are 
simultaneously excited through interfering amplitudes \cite{BlattBiedenharn} - 
this is observed in $\mathrm{\pi^- p \to K^0\Lambda}$ scattering \cite{knasel75,baker78}.
Hence, this distribution in the $\mathrm{K^+\Lambda}$ Jackson frame is 
a strong indication
in the present analysis of one-dimensional distributions that more than one $N^*$ resonance 
with opposite parity participates in the production process, namely $\mathrm{N(1650)S}_{11}$, $\mathrm{N(1710)P}_{11}$, 
and $\mathrm{N(1720)P}_{13}$. This section is concluded by stressing that this
finding is fully consistent with that extracted from the Dalitz plot \cite{toflambda06} 
which describes the correlations of the exit channel only. 

\subsubsection{Angular distributions in helicity frames}
\label{doppeltdiff}
In a reaction of type $ab\to$ {\it 123} the helicity angle in a respective 
helicity frame\footnote{For reactions of type $ab\to \it 123$ 
the helicity frame is defined as the
Lorentzian frame in which the center of mass of the particles {\it (2,3)} is at rest ($\vec{p}_3=-\vec{p}_2$), 
{\it i.e.} it is the identical Lorentzian frame as the respective Jackson frame.
The choice of the word ``frame'' relates to 
the reference axis, which in case of the helicity frame is the direction of particle {\it 1}.
As in the case of Jackson frames, three helicity frames can be constructed by cyclic permutation for the 
three-body final state ({\it {R23, R31, R12}}).
Nomenclature: $\theta^\mathrm{R23}_{13}$ is the angle of particle {\it 3} with respect to the reference axis {\it 1} (subscript), 
measured in the rest frame of particle {\it 2} and {\it 3} (superscript).} 
interrelates the three particles of the exit channel; in fact, 
the helicity angle distribution is simply a special type of projection of a Dalitz plot.
A uniformly populated Dalitz plot corresponds to isotropic helicity angle distributions whereas
all physical and/or kinematical effects distorting the Dalitz plot must result in characteristic 
distributions in helicity frames.
For example, an isolated narrow resonance decaying into the ({\it 23}) system manifests itself as a ``band''
extending along the $m^2_{13}$ (or likewise $m^2_{12}$) axis in a Dalitz plot. 
The properties of mass and width of the resonance are seen in the 
({\it 12}) and ({\it 13}) helicity frames while the decay pattern characteristic for the angular 
momentum of the resonance shows up in the ({\it 23}) helicity frame. 

Final state interaction also distorts a Dalitz plot and consequently FSI effects are also seen in helicity angle distributions. 
A strong FSI, for example,  between particle {\it 1} and {\it 3} leads 
to an enhancement at $\mathrm{cos}\theta^\mathrm{R23}_{13}=1$ and $\mathrm{cos}\theta^\mathrm{R12}_{32}=-1$.
If, however, FSI-effects and various (resonant and non-resonant) reaction mechanisms contribute
the situation becomes by far more complicated and quantitative conclusions can only be drawn with caution.
Then, theoretical models have to treat all contributions in a
coherent manner and their results have to be confronted with the data.
Such an approach was adopted for the $\mathrm{pp \to pK^+\Lambda}$ channel 
and an excess energy of 130 MeV by Sibirtsev {\it et al.} \cite{sibi06b}.
These authors pointed out that helicity distributions are ideally suited to determine the various contributions 
to the reaction mechanism in a quantitative manner.

The angular distributions in the three helicity frames
are shown in the lower rows of figs.~\ref{fig:2950lambda} to \ref{fig:3200lambda}.
All distributions deviate significantly from isotropy. The dashed histograms, which are isotropic, 
again are the results of Monte Carlo data filtered with the experimental distribution as measured in 
the $\mathrm{p\Lambda}$-Jackson frame. The filter is without any effect on the helicity angle as the latter is solely 
an exit channel property.
The parameters of Legendre polynomial fittings to the experimental distributions are shown
in tables \ref{tab:fitparasL1} to \ref{tab:fitparasL3}.

\begin{figure}[bbb]
\resizebox{0.5\textwidth}{!}{ \includegraphics{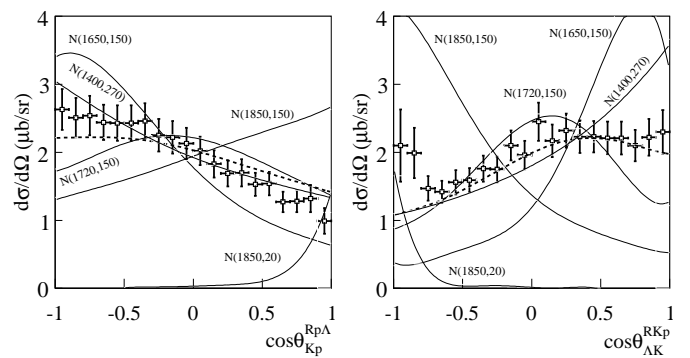} }
\caption{Angular distributions  measured in the $\mathrm{p\Lambda}$ and $\mathrm{K^+p}$ helicity frames.
The lines are the Monte Carlo result for various nucleon resonances ($N(m,\Gamma)$). The dashed line is the result 
for a incoherent sum of Monte Carlo data (see text). }
\label{fig:finalbild}  
\end{figure}

As mentioned above, the distributions in helicity fra\-mes both depend on and reflect, via kinematics,
the masses and the widths of participating nucleon resonances.
In order to stress these aspects, the experimental data taken at $\epsilon=239\,\mathrm{MeV}$  are shown  
in fig.~\ref{fig:finalbild} again, left hand side in the $\mathrm{p\Lambda}$, 
right hand side in the $\mathrm{K^+p}$ helicity frames. 
In the same pictures are included the results of Monte Carlo simulations of single 
nucleon resonances ($N^*\to\mathrm{K^+\Lambda}$) with mass $m$ and width $\Gamma$ abbreviated as $N(m,\Gamma)$.
In addition, the result for an incoherent sum of three Monte Carlo data-sets 
(phase space, $\mathrm{N(1650,150)}$, 
$\mathrm{N(1720,150)}$ - all entering with equal weight) is shown as a dashed line.
Some general conclusions will now be drawn by comparing these Monte Carlo distributions with the measured ones.

The effect of narrow resonances (N(1600), N(1650), N(1720), N(1850); $\Gamma=20\,\mathrm{MeV}$) 
is studied first; they were found to cause strongly localized distortions of the distributions
in both helicity frames. As an example the result for $\mathrm{N(1850,20)}$, assumed to have a cross section of
$2.5\,\mathrm{\mu b}$, is shown in fig.~\ref{fig:finalbild}.
In this case the distribution in the $\mathrm{K^+p}$ helicity frame would
leave room for such a resonance to contribute, however the corresponding characteristic distortion in the 
$\mathrm{p\Lambda}$ helicity frame is not observed experimentally. 
The results obtained for all the other narrow resonances are incompatible with the data in both frames.
Hence, any hypothetical narrow resonance (``missing'' or exotic) is excluded to contribute 
on a level of about $2.5\,\mathrm{\mu b}$. 

Resonances  with masses $m_{N^*} \geq 1850\,\mathrm{MeV}$ and various widths of  $\Gamma$=150 to $400\,\mathrm{MeV}$
yield distributions incompatible with the measured ones.
As an example for this class of resonances, the effect of the $\mathrm{N(1850,150)}$
is shown in fig.~\ref{fig:finalbild}. The distributions of measured and Monte Carlo data 
clearly show opposite slopes in either frame.
Thus a whole set of resonances, namely $\mathrm{N(1900)P_{13}}$ \cite{pdg06},
$\mathrm{N(1840)P}_{11}$ \cite{sarantsev05}, $\mathrm{N(1875)D}_{13}$ \cite{sarantsev05}, 
$\mathrm{N(1895)D}_{13}$ \cite{McNabb04,glander03}, and $\mathrm{N(1950)D}_{13}$ \cite{pennerNmosel02} 
is unlikely to contribute to the reaction $\mathrm{pp\to pK^+\Lambda}$. 

The kinematical effect of a light nucleon resonance was studied by means of a simulation of an $\mathrm{N(1400,270)}$ resonance. 
As its central mass is smaller than the sum of the masses of kaon and $\mathrm{\Lambda}$ the production process can only proceed
via the broad tail of the resonance. The numerical values for mass and width 
are chosen according to the result of the resonance isobar model of ref.\cite{luizou06} which 
explains well the energy dependence of the ratio $R_{\mathrm{\Lambda}/\Sigma^0}$ by invoking only this resonance. 
The distributions generated by this single resonance
follow fairly well the experimental one in the $\mathrm{p\Lambda}$ but fail 
to produce that observed in the $\mathrm{Kp}$  helicity frame. 
Hence, a broad sub-threshold resonance alone is not able to account for the 
kinematic situation found in the $\mathrm{pK^+\Lambda}$ final state.

The same observation holds true for both N(1650,150) and 
N(1720,150) which on their own are not able to describe the 
observed angular distributions in fig.~\ref{fig:finalbild}. Similarly, a distribution 
according to pure three-body phase does not follow the data as it leads to isotropic
distributions (not shown). On the other hand, 
the sum of these two resonances and phase space (represented by the dashed line)
follows the experimental data very well.
Of course, this mere incoherent superposition reflects only a particular kinematic situation (omitting FSI effects) and
does not describe the dynamics of the $\mathrm{pK^+\Lambda}$ production process. Therefore, beyond kinematics, it should be interpreted with caution, especially 
as the analysis of distributions in the $\mathrm{K^+\Lambda}$ Jackson frame strongly indicated  interference effects 
due to resonances with certain spins and opposite parity, namely
$\mathrm{S_{11}}$, $\mathrm{P_{11}}$, and $\mathrm{P_{13}}$. These are just the states $\mathrm{N(1650)S}_{11}$, $\mathrm{N(1710)P}_{11}$/$\mathrm{N(1720)P}_{13}$
which were included in the Monte Carlo simulation. 
 Thus, the very good reproduction of the experimental data by a curve resulting from
an incoherent sum of three Monte Carlo data sets 
supports, from a kinematical point of view, that the resonances involved are, indeed, 
$\mathrm{N(1650)S}_{11}$, $\mathrm{N(1710)P}_{11}$/ $\mathrm{N(1720)P}_{13}$.
 
Finally, the implications of the $\mathrm{p\Lambda}$-helicity angle distributions for the 
explanation of the $\mathrm{\Lambda}$ to $\mathrm{\Sigma^0}$ cross section ratio near the threshold are discussed.
In \cite{deswat63} $R_{\mathrm{\Lambda/\Sigma^0}}=27$ was predicted from the ratio of the $KNY$ coupling constants 
$g^{2}_{N\Lambda K}/g^{2}_{N\Sigma K}$. Obviously, for this prediction to apply the production process must be dominated by 
kaon exchange (fig.~\ref{fig:ractionmechnism}a).
Pure kaon exchange, however, should lead to an isotropic distribution in the $\mathrm{p\Lambda}$-helicity frame \cite{sibi06b}, 
which is not observed experimentally.
Hence, kaon exchange cannot be the dominant production process and nucleon resonances 
have also to be taken into account at the excess energies considered here.
Then it is reasonable to assume that resonant contributions are also present near the threshold,
as, e.g., the $\mathrm{N(1650)S}_{11}$ resonance (width $\approx 165\,\mathrm{MeV}$ \cite{pdg06}) can easily 
be produced at threshold energies. Hence, the explanation of the $R_{\mathrm{\Lambda/\Sigma^0}}$  
to be solely related to the $NKY$ coupling constants is not likely to be correct.

The angular distributions in the $\mathrm{K^+\Lambda}$  helicity frame are shown in the lower-right frames 
of figs.~\ref{fig:2950lambda} to \ref{fig:3200lambda}. 
An apparent enhancement towards $\mathrm{cos}\theta^{\mathrm{RK\Lambda}}_{\mathrm{p\Lambda}}=1$
is observed for all excess energies.
This asymmetry cannot be caused by any single nucleon resonance, as in this case the 
distribution would be either isotropic (decay angular momentum $L=0$)
or symmetric with respect to $\mathrm{cos}\theta=0$ ($L\ge1$).

The $\mathrm{p\Lambda}$ FSI introduces an asymmetry with enhanced cross section 
towards $\mathrm{cos}\theta^{\mathrm{RK\Lambda}}_{\mathrm{p\Lambda}}=1$.
However, FSI effects are limited to relative energies of the $\mathrm{p\Lambda}$ system of just a few MeV and 
therefore influence only a limited region of the Dalitz plot at the excess energies considered here. 
Thus, it is difficult to see how the observed asymmetric distribution in the $\mathrm{K^+\Lambda}$ helicity frame,
which is a projection of the whole Dalitz plot, can be explained solely by FSI.
If one excludes exotic resonances decaying into the $\mathrm{p\Lambda}$ or $\mathrm{pK^+}$ channel,
which would lead via kinematics to an asymmetry in the $\mathrm{K^+\Lambda}$ helicity frame, one is left with the conclusion that 
a coherent interference of various $N^*$ resonances (including FSI) dominates the reaction.
It is known that the interference of 
resonances of different parity leads to asymmetric angular distributions \cite{BlattBiedenharn}.
The analysis of the distributions in the $\mathrm{K^+\Lambda}$ Jackson frame evidenced that  more than one resonance 
of opposite parity are involved. The
discussion of fig.~\ref{fig:finalbild} showed that a combined contribution of the
$\mathrm{N(1650)S}_{11}$ and $\mathrm{N(1710)P}_{11}$/$\mathrm{N(1720)P}_{13}$, which have opposite parity, 
is kinematically supported.
Hence, the mentioned interference of resonances with different parity is possible and likely to cause the asymmetry 
observed in the $\mathrm{K^+\Lambda}$  helicity frame. 

We like to end this section with a comment on a very recent observation of an exotic resonant state $X$ 
($m_{\mathrm{X}}$$=2267\, \mathrm{MeV/c^2}$, $\Gamma_{\mathrm{X}}$ = 118 MeV/c$^2$) which decays into
two baryons, a 
non-strange proton and a strange $\mathrm{\Lambda}$ hyperon \cite{DISTO2010}. 
The resonance is interpreted to be a deeply bound, compact $\mathrm{K^-pp}$ cluster (binding energy $\approx$ 100 MeV)
which, hence, could be a possible gateway towards cold and dense kaonic nuclear matter. 
The measurement was carried out with a beam momentum of 3670 MeV/c which corresponds to an excess energy
of $\epsilon=\sqrt{s}-m_{K^+}-m_X=221$ MeV. At this energy, the
contribution of this exotic $\mathrm{K^+}X$ production to the total $\mathrm{pp\to pK^+\Lambda}$ cross section
is found to be about 20\%.

Compared to \cite{DISTO2010} the beam momenta considered in this paper (2950, 3059, 3200 MeV/$c$) 
correspond to considerably smaller excess energies of -9, 26, and 71 MeV for $\mathrm{K^+}X$ production.
Due to the width of the resonance the $\mathrm{K^+}X$ final state is kinematically accessible even at the 
negative excess energy of -9 MeV. 
However, all three excess energies are smaller than the width of the resonance. Thus, the 
strength of possible contributions must consequently be considerably smaller compared to \cite{DISTO2010}.
Contributing only on a few percent level, such a weak signal is obviously difficult
to observe directly in figs.~\ref{fig:2950lambda}-\ref{fig:finalbild}.
Theory is now asked to explore the benefit of adding this exotic state, 
in addition to ``standard'' nucleon resonances decaying into $\mathrm{K^+\Lambda}$, when reproducing our data.

\begin{figure}[htt]
\resizebox{0.5\textwidth}{!}{ \includegraphics{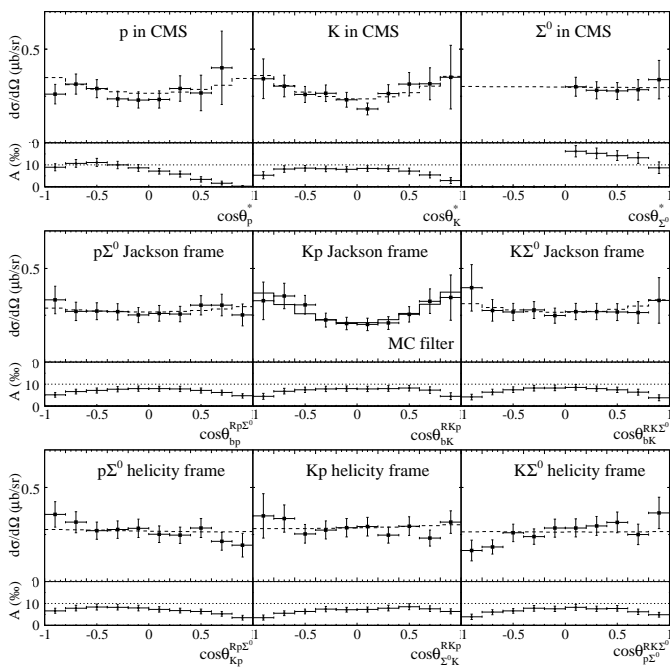} }
\caption{Angular distributions of the particles in the overall CMS, Jackson and  helicity frames (top-down) 
for the reaction $\mathrm{pp \to pK^+\Sigma^0}$ measured at an excess energy of $\epsilon=162$ MeV 
($p_{beam}=3059\, \mathrm{MeV/c}$). Error bars for each data point are the square root of the quadratic sum 
of the statistical, acceptance, and 
signal-to-background-separation uncertainty.
The solid histogram in the Kp Jackson frame represents the Legendre polynomial 
of table \ref{tab:fitparasS} which is used as MC filter. Its effect on all other angular distributions is shown
by the dashed histograms. Below each angular distributions the differential acceptance is shown. 
}
\label{fig:3059sigma}  
\end{figure}

\subsection{Differential distributions: $\mathrm{pp}\to \mathrm{pK^+\Sigma^0}$}
The data set taken at $p_{beam}=3059\,\mathrm{MeV/c}$ ($\epsilon_\mathrm{\Sigma^0}=162$ 
$\mathrm{MeV}$) provides for the first time direct insight into 
the reaction dynamics of proton-proton induced $\mathrm{pK^+\Sigma^0}$ production 
as differential observables are accessible. 
These angular distributions are shown in fig.~\ref{fig:3059sigma}; 
the coefficients from the least square fitting are listed in table \ref{tab:fitparasS}.
The histograms represent filtered Monte Carlo data, which will be explained in the course of the discussion. 
The arrangement within the figure is analogous to the case of $\mathrm{\Lambda}$ production 
(figs.~\ref{fig:2950lambda} to \ref{fig:3200lambda}), {\it i.e.} the first row shows 
distributions in the overall CMS while in the second and third row distributions in the Jackson and helicity frames are
presented, respectively. In order to avoid repetition, we will draw on arguments presented above in order to directly
deduce conclusions on $\mathrm{pK^+\Sigma^0}$ production.
Emphasis will be put on a comparison of both hyperon production channels closest in excess energy where, for the time being, the
difference in $\epsilon_\mathrm{\Lambda}=204\,\mathrm{MeV}$  
and  
$\epsilon_\mathrm{\Sigma^0}=162\,\mathrm{MeV}$ 
is ignored. This seems justified
as the reaction mechanism for $\mathrm{\Lambda}$ production was found, as shown above, not to change dramatically with excess energy.

\begin{table}[ttt]
\caption{Coefficients of Legendre polynomials (in units of $\mu $b/sr) determined by least square fitting to angular distributions 
of the reaction $\mathrm{pp\to pK^+\Sigma^0}$ at
$\epsilon=162\,\mathrm{MeV}$ ($p_{beam}=3059\,\mathrm{MeV/c}$).} 
\label{tab:fitparasS}
     	\begin{tabular}{@{}lrrr}
  		\hline\noalign{\smallskip}
 cos & $a_0\;\;\;\;\;$ & $a_1\;\;\;\;\;$ & $a_2\;\;\;\;\;$   \\   		\noalign{\smallskip}\hline\noalign{\smallskip}
$\theta^*_{p}$ 						&  $ 0.28 \pm 0.04 $ &   $ 0.03 \pm 0.07 $ &   $ 0.07 \pm 0.08 $  \\ \noalign{\smallskip}							
$\theta^*_{K}$ 						&  $ 0.29 \pm 0.02 $ &   $ 0.01 \pm 0.05 $ &   $ 0.14 \pm 0.06 $  \\ \noalign{\smallskip}							
$\theta^*_{\Sigma^0}$ 				&  $ 0.29 \pm 0.03 $ &   ---$\;\;\;\;\;\,$ &  $ 0.01 \pm 0.07 $  \\ \noalign{\smallskip}\hline\noalign{\smallskip}  
$\theta^{Rp\Sigma^0}_{bp}$ 			&  $ 0.27 \pm 0.02 $ &   $ -0.00 \pm 0.03 $ &	$ 0.03 \pm 0.04 $ \\ \noalign{\smallskip}							
$\theta^{RKp}_{bK}$ 				&  $ 0.28 \pm 0.02 $ &   $ -0.02 \pm 0.04 $ &	$ 0.16 \pm 0.05 $ \\ \noalign{\smallskip}							
$\theta^{RK\Sigma^0}_{bK}$ 			&  $ 0.28 \pm 0.02 $ &  $ -0.01 \pm 0.04 $ &   $ 0.04 \pm 0.05 $  \\ \noalign{\smallskip}\hline\noalign{\smallskip} 
$\theta^{Rp\Sigma^0}_{Kp}$ 			&  $ 0.27 \pm 0.02 $ &  $ -0.06 \pm 0.03 $ &   $ 0.00 \pm 0.04 $  \\ \noalign{\smallskip}							
$\theta^{RKp}_{\Sigma^0 K}$ 		&  $ 0.28 \pm 0.02 $ &  $ -0.03 \pm 0.04 $ &   $ 0.03 \pm 0.05 $  \\ \noalign{\smallskip}							
$\theta^{RK\Sigma^0}_{p\Sigma^0}$ 	&  $ 0.26 \pm 0.02 $ &   $ 0.07 \pm 0.03 $ &  $ -0.04 \pm 0.04 $  \\ \noalign{\smallskip}							
  	\noalign{\smallskip}\hline
	\end{tabular}
\end{table}

The distributions in the overall CMS and the Jackson frames for $\mathrm{\Lambda}$ and $\mathrm{\Sigma^0}$ 
production (first/second row in 
figs.~\ref{fig:2950lambda} and \ref{fig:3059sigma}) 
not only differ apparently but in addition show an opposite behavior.
An inspection of the parameters given in table \ref{tab:fitparasS} 
shows that
the proton and $\mathrm{\Sigma^0}$ CMS distributions as well as the those in the 
$\mathrm{p\Sigma^0}$ and $\mathrm{K^+\Sigma^0}$ Jackson frame
are compatible with isotropy whereas the corresponding distributions in the 
$\mathrm{pK\Lambda}$ channel (table \ref{tab:fitparasL1}) 
show strong anisotropy; 
for the $\mathrm{pK\Lambda}$ channel the kaon distribution in the CMS and that in the $\mathrm{K^+p}$ Jackson frame
feature a pronounced anisotropy whereas the corresponding distributions in the $\mathrm{\Lambda}$ channel show
strong isotropy.
These finding are the first proof that
the reaction mechanisms differ significantly for the reaction 
$\mathrm{pp\to pK^+\Lambda}$ and $\mathrm{pp\to pK^+\Sigma^0}$.

From the CMS distributions one directly concludes that 
the $\mathrm{K^+ - (p\Sigma^0)}$ system shows some indication of $L=1$
while the $\mathrm{\Sigma^0}$ hyperon 
has no relative angular momentum with respect
to the $\mathrm{pK^+}$ subsystem.
This is in line with the mainly isotropic distribution observed in the $\mathrm{p\Sigma^0}$ 
Jackson frame. Hence, if intermediate resonances are involved ($N^*$ {\it or }  
$\Delta^*$), the relative angular momentum shared by the $\mathrm{p}-N^*$ or $\mathrm{p}-\Delta^*$ system is 
$L=0$, remarkably different from the case of $\Lambda$ production. 

The fact that no angular momentum is involved in the $\mathrm{p}-N^*/\Delta^*$ system
signifies that only particular partial waves in 
the entrance channel can participate. The nuclear reaction theory developed 
by Blatt and Biedenharn \cite{BlattBiedenharn} shows that isotropic angular distributions are only possible if one 
of the three following quantities is zero: entrance or exit channel angular momentum or total angular 
momentum. This condition is always fulfilled for the entrance channel states $\mathrm{^{1}S_0}$ and 
$\mathrm{^{3}P_0}$, irrespectively of the final state involved. 
It is also met for $\mathrm{^{3}P_1}$ if it produces $\mathrm{S_{11}}$ or $\mathrm{S_{31}}$ resonances in the exit channel.
Thus, the observed isotropy significantly constrains partial waves being possible in entrance and exit channel.
A conclusion like that could not be drawn 
in the case of the $\mathrm{pp \to pK^+\Lambda}$ reaction as the corresponding angular 
distributions are anisotropic.

The distribution in the $\mathrm{K^+\Sigma^0}$ Jackson frame shows isotropy within uncertainty.
Hence, there is only little room for resonances with intrinsic angular momentum. 
Thus, from the list of nucleon resonances with assumed decay branches to $\mathrm{K^+\Sigma^0}$ \cite{pdg06},
the resonances to be possibly involved are most likely
$\mathrm{N(1650)S}_{11}$, and $\mathrm{\Delta(1900)S}_{31}$  
while $\mathrm{N(1710)P}_{11}$, $\mathrm{N(1720)P}_{13}$, and $\mathrm{\Delta(1750)P}_{31}$ seem less probable to contribute.

The distribution in the $\mathrm{K^+p}$ Jack\-son frame shows an\-isotropy. This indicates that
the proton-kaon subsystem carries a relative angular momentum of $L=1$. 
This cannot be the decay angular momentum of $N^*$ or $\Delta^*$ resonances as it is impossible for them to disintegrate
into a non-strange proton and a strange kaon - only an as yet unknown pentaquark state could feature such a decay. 
The anisotropy observed is  rather conjectured to be a strong indication of the presence of kaon exchange 
(fig.~\ref{fig:ractionmechnism}a).
This again discriminates between the $\mathrm{\Sigma^0}$ and the $\mathrm{\Lambda}$ channel.

As in the case of the  $\mathrm{\Lambda}$ production a filter was set 
on those distributions which deviate the most from isotropy.
It is found that the filter on the distribution in the Kp Jackson frame (solid line histogram) is well suited to 
simultaneously describe the data in the CMS and Jackson frames (dashed line histograms).

The distributions in the three helicity frames are presented in the bottom row of 
fig.~\ref{fig:3059sigma}. 
Compared to those for the $\mathrm{\Lambda}$ production shown 
in fig.~\ref{fig:2950lambda} ($\epsilon=204\,\mathrm{MeV}$) 
these distributions feature a smaller, however, still significant anisotropy. 
In particular, the asymmetric distribution in the $\mathrm{K^+\Sigma^0}$ helicity frame show
interference effects which, in the present case of $\Sigma^0$ production, may involve  
nucleon as well as $\Delta$ resonances, kaon exchange potentially carrying angular momentum as well as 
all three possible final state interactions.

\subsection{Summary}
Associated strangeness production was investigated in the reactions $\mathrm{pp\to pK^+\Lambda}$ 
and $\mathrm{pp\to pK^+\Sigma^0}$ 
using data collected by the time-of-flight spectrometer {\it COSY-TOF}.
Data were analyzed for three different beam momenta ($p_{beam}=2950, 3059, 3200\, \mathrm{MeV/c}$), 
which correspond  
to excess energies of 204, 239, and $284\,\mathrm{MeV}$ 
in the case of $\mathrm{\Lambda}$ production whereas the neutral $\mathrm{\Sigma}$ is produced  127, 162, 
and $207\,\mathrm{MeV}$ above the threshold.

These measurements extend the experimental knowledge of both reactions by roughly $\mathrm{80\,MeV}$ into an energy range 
formerly believed to show an sizable enhancement of the ratio of 
the $\mathrm{\Lambda}$ to $\mathrm{\Sigma^0}$ cross section. 
At $\epsilon=204\,\mathrm{MeV}$ this ratio is found to be  $2.5\pm 0.4$ and shows the 
high energy limit, so far measured for $\epsilon>700\,\mathrm{MeV}$, to be reached already in the region of the moderate
excess energies considered here.

In the case of $\mathrm{pp\to pK^+\Lambda}$ differential cross sections were obtained in the CM-, 
Jackson-, and helicity-frames for all three excess energies.
Strong evidence  was found for a production scenario including intermediate nucleon resonan\-ces. Especially
a process which involves $\mathrm{N(1650)S_{11}}$, $\mathrm{N(1710)P_{11}}$, and/or $\mathrm{N(1720)P_{13}}$ 
is deduced from the data; 
$\mathrm{D_{13}}$, $\mathrm{D_{15}}$, and $\mathrm{F_{15}}$ resonances are unlikely to contribute.
Kaon exchange with angular momentum $L\geq1$ is excluded while $L=0$ may be present, however, 
surely not being the dominant process. 

For the reaction $\mathrm{pp}\to\mathrm{pK^+\Sigma^0}$ the first differential 
data have been shown ($\epsilon=162\,\mathrm{MeV}$).
The distributions in the overall CMS as well as in the various Jackson frames feature
an opposite behavior with respect to those observed for $\mathrm{\Lambda}$ production.
Only entrance channel partial waves $\mathrm{^{1}S_0}$, $\mathrm{^{3}P_0}$, and $\mathrm{^{3}P_1}$ 
are found to be involved. Strong indication is found for kaon exchange carrying angular momentum. 
If resonances contribute to $\mathrm{\Sigma^0}$ production they are likely to be 
of the type $\mathrm{S_{11}}$ and $\mathrm{S_{31}}$.

Thus, one of the key results of this paper is the proof that the
reaction mechanisms for $\mathrm{\Sigma^0}$ and $\mathrm{\Lambda}$ production differ decisively. 
With this differential data at hand theory is now challenged to advance a model which 
simultaneously describes the data presented.

\section*{Acknowledgment}
The authors would like to express their gratitude to the COSY staff for the operation 
of the accelerator during the experiments. Discussions with V.~A.~Nikonov, A.~V.~Sarantsev and A.~A.~Sibirtsev 
are gratefully acknowledged.
This work was supported in part by grants from BMBF and Forschungszentrum J\"ulich (COSY-FFE).


%

\begin{thebibliography}{10}
%
%

\bibitem{grzonka97}
D.~Grzonka and K.~Kilian,
{Nucl.~Phys.~A} {\bf 626}, {41C-54C} {(1997)}

\bibitem{balewski97}
J.~T.~Balewski  {\it et al.},
{Nucl.~Phys.~A} {\bf 626}, {85C-92C} {(1997)}

\bibitem{balewski98}
J.~T.~Balewski  {\it et al.},
{Phys.~Lett.~B} {\bf 420}, {211-216} {(1998)}

\bibitem{sewerin99}
S.~Severin {\it et al.}, 
{Phys.~Rev.~Lett.} {\bf 83}, {682} {(1999)}

\bibitem{COSY11kowina04}
P.~Kowina {\it et al.},
{Eur.~Phys.~J.~A} {\bf 22}, {293-299} {(2004)}

\bibitem{toflambda06}
M.~Abd El-Samad {\it et al.},
{Phys.~Lett.~B} {\bf 632}, {27-34} {(2006)}

\bibitem{valdau07}
 Yu.~Valdau {\it et al.},
{Phys.~Lett.B} {\bf 652}, {245-249} {(2007)} (e-Print: nucl-ex/0703044)

\bibitem{baldini88}
A.~Baldini {\it et al.},
Landolt-B\"ornstein, New Series, I/12b (1988)

\bibitem{deswat63}
J.~J.~De Swart, 
Rev.~Mod.~Phys.~{\bf 35}, 916 (1963)

\bibitem{sibi06}
 A.~A.~Sibirtsev  {\it et al.}, 
{Eur.~Phys.~J.~A} {\bf 29}, {363} {(2006)}

\bibitem{shyam06}
R.~Shyam, 
{Phys.~Rev.~C} {\bf 73}, {035211} {(2006)} [nucl-th/0512007]

\bibitem{luizou06}
B.~C.~Lui and B.~S.~Zou, 
{Phys.~Rev.~Lett.} {\bf 96}, {042002} {(2006)}

\bibitem{gaspa00}
 A.~Gasparian {\it et al.}, 
{Phys.~Lett.~B} {\bf 480} {273} (2000)

\bibitem{dillig06}
M.~Dillig and M.~Schott, 
arXiv: nucl-th/0604059v1

\bibitem{saturneandfriends1}
R.~Frascaria {\it et al.}, 
{Nuovo Cimento} {\bf 102A}, {561} {(1989)},

\bibitem{saturneandfriends2}
R.~Siebert, 
{Nucl.~Phys. A} {\bf 567} {819} {(1994)}

\bibitem{laget01Nlaget91}
J.-M.~Laget, 
{Phys.~Lett.~B} {\bf 259} {24} {(1991)},
{Nucl.~Phys.~A} {\bf 691}, {11c} {(2001)}

\bibitem{capstickNroberts98}
S.~Capstick and W.~Roberts,
{Phys.~Rev.~D} {\bf 58}, {074011} {(1998)}

\bibitem{bradford06}
 R.~Bradford {\it et al.},
{Phys.~Rev.~C.} {\bf73}, {035202} {(2006)}

\bibitem{sumihama06}
M.~Sumihama  {\it et al.},
{Phys.~Rev.~C} {\bf73}, {035214} {(2006)}

\bibitem{hicks07}
K.~Hicks {\it et al.},
{Phys.~Rev.~C} {\bf 76}, {042201(R)} {(2007)}

\bibitem{mart06}
T.~Mart and A.~Sulaksono,
{Phys.~Rev.~C} {\bf 74}, {055203} {(2006)}

\bibitem{tram98}
M.Q.~Tran {\it et al.},
{Phys.Lett.~B} {\bf 445}, {20} {(1998)}

\bibitem{pdg06}
PDG06, Journal of Phys.~G, {\bf 33}, 1-1232 (2006)

\bibitem{sarantsev05}
A.V.~Sarantsev {\it et al.},
{Eur.~Phys.~J.~A} {\bf 25}, {441} {(2005)}

\bibitem{McNabb04}
J.W.C.~McNabb {\it et al.},
{Phys.Rev.~C} {\bf 69}, {042201(R)} {(2004)}

\bibitem{glander03}
K.-H.~Glander {\it et al.},
{Eur.~Phys.~J.~A} {\bf 19}, {251} {(2004)}

\bibitem{pennerNmosel02}
G.~Penner and U.~Mosel,
{Phys.~Rev.~C} {\bf 66}, {055212} {(2002)}

\bibitem{capstickNroberts2000}
{S.~Capstick and C.~Roberts},
{Prog.~Part.~Nucl.~Phys} {\bf 45}, {2541} {(2000)}

\bibitem{eyrich07}
W.~Eyrich,
{Eur.~Phys.~J.~A} {\bf 31}, {435-440} {(2007)}

\bibitem{schoeder07}
 W.~Schroeder for the {\it COSY-TOF} collaboration,
{Eur.~Phys.~J.~A} {\bf 31}, {503-505} {(2007)}

\bibitem{tofthetaplusdead} M.~Abdel-Bary  {\it et al.} ({\it COSY-TOF} collaboration),
 Phys.~Lett.~B {\bf 649}, 252 (2007)

\bibitem{tofbeschreibung}
M.~Dahmen, PhD Thesis, Rheinische Friedrich-Wilhelm-Universit\"at Bonn, J\"ul-3140, ISSN 0944-2952 (1995).

\bibitem{Erlangenstart}
R.~Bilger  {\it et al.},
{Phys.~Lett.~B} {\bf 420}, {217-224} {(1998)}

\bibitem{target} A.~Hassan {\it et al.},
 Nucl.~Instrum.~Methods Phys. Res. A {\bf 425}, 403 (1999)

\bibitem{forewardhodo} 
M.~Dahmen  {\it et al.},
 Nucl. Instruments Methods Phys. Res. A {\bf 348}, 97 (1994)

\bibitem{barrel} A.~B{\"o}hm {\it et al.}, 
 Nucl.~Instruments~Methods~Phys.~Res.~A {\bf 443}, 238 (2000)

\bibitem{lasvegasbeschreibungBrand}
S.~Brand, PhD Thesis, Ruhr-Universit\"at Bochum (1995)

\bibitem{lasvegasbeschreibungZielinsky}
U.~Zielinsky, PhD Thesis, Ruhr-Universit\"at Bochum (1999)

\bibitem{edda00}
D.~Albers  {\it et al.},
{Phys.~Rev.~Lett.} {\bf 78}, {1652} {(1997)}

\bibitem{MSWdiss}
{M.~Schulte-Wissermann},
PhD Thesis, TU Dresden (2004)

\bibitem{WU07}
{M.~Abdel-Bary\it et al.}, 
{Physics Letters B} {\bf 662}, {14-18} {(2008)}

\bibitem{mswpaper06}
M.~Abdel-Bary {\it et al.}, 
{Phys.Lett.~B} {\bf 647}, {351-357} {(2007)}

\bibitem{wupaper09}
M.~Abdel-Bary {\it et al.}, 
Accepted by Eur.~Phys.~J.~A (2010)\newline
arXiv:1001.3043v1 [nucl-ex]

\bibitem{dietrich06}
 J.~Dietrich,
Diploma Thesis, TU Dresden, 2006

\bibitem{erlangenzukunft10}
M.~Abdel-Bary {\it et al.}, 
Phys.~Lett. {\bf B} (in print)

\bibitem{fickinger62}
W.J.~Fickinger {\it et al.},
{PR} {\bf 125}, {2082} {(1962)}

\bibitem{Byckling}
{E.~Byckling and K.~Kajantie},
{\it Particle Kinematics}, 
{John Wiley \& Sons} {(1973)}

\bibitem{GJ}
K.~Gottfried and J.~D.~Jackson,
{Nuovo Cimento} {\bf 33}, {309} {(1964)}

\bibitem{faeldtwilkin97}
G.~F\"aldt and C.~Wilkin,
{Z.~Phys.} {\bf A357}, {241} {(1997)}

\bibitem{knasel75}
T.~M.~Knasel {\it et al.},
{Phys.~Rev.~D} {\bf 11}, {1} {(1975)}

\bibitem{baker78}
R.~D.~Baker {\it et al.},
{Nucl.~Phys.~B} {\bf 141}, {29} {(1978)}

\bibitem{balestra99}
F.~Balestra {\it et al.},
{Phys.~Rev.~Lett.} {\bf 83}, {1534} {(1999)}

\bibitem{BlattBiedenharn}
 I.~M.~Blatt and L.~C.~Biedenharn,
{Rev.~Mod.~Phys.} {\bf 24}, {258} {(1952)}

\bibitem{DISTO2010}
 T.~Yamazaki {\it et al.}, 
{Phys.~Rev.~Lett.} {\bf 104}, {13250-1} {(2010)}

\bibitem{sibi06b}
 A.~A.~Sibirtsev {\it et al.}, 
{Eur.~Phys.~J.~A} {\bf 27}, {269-285} {(2006)}

\end{thebibliography}
%

\appendix
\section{Datatables}

\begin{table*}[hhh]
\caption{Cross sections in $\mathrm{\mu b/sr}$ for the reaction $\mathrm{pp\to pK^+\Lambda}$, $p_{beam} = 2950\,\mathrm{MeV/c}$, $\epsilon = 204\,\mathrm{MeV}$, fig.~\ref{fig:2950lambda}.
} 
\label{tab:diffobs2950KpL}
     	\begin{tabular}{@{}rccccccccc}
  		\hline\noalign{\smallskip}
 cos$\,\theta$ & $\frac{d\sigma}{d\Omega}(\theta^*_\mathrm{p})$$\;$ & $\frac{d\sigma}{d\Omega}(\theta^*_\mathrm{K^+})$$\;$ & $\frac{d\sigma}{d\Omega}(\theta^*_\mathrm{\Lambda})$$\;$ & $\frac{d\sigma}{d\Omega}(\theta^{Rp\Lambda}_{bp})$$\;$ & $\frac{d\sigma}{d\Omega}(\theta^{RKp}_{bK})$$\;$ & $\frac{d\sigma}{d\Omega}(\theta^{RK\Lambda}_{bK})$$\;$ & $\frac{d\sigma}{d\Omega}(\theta^{Rp\Lambda}_{Kp})$$\;$ & $\frac{d\sigma}{d\Omega}(\theta^{RKp}_{\Lambda K})$$\;$ &  $\frac{d\sigma}{d\Omega}(\theta^{RK\Lambda}_{p\Lambda})$$\;$  \\ 
          \noalign{\smallskip}\hline\noalign{\smallskip}
  -0.95   &   $ 3.88 \pm 0.52 $ & $ 2.46 \pm 1.58 $ &        ---         & $ 4.25 \pm 0.65 $ & $ 2.34 \pm 0.63 $ & $ 2.75 \pm 0.70 $   & $ 1.56 \pm 0.22 $ & $ 2.40 \pm 0.71 $ & $ 1.76 \pm 0.75 $     \\ 
  -0.85   &   $ 2.58 \pm 0.25 $ & $ 2.04 \pm 0.50 $ &        ---         & $ 3.00 \pm 0.31 $ & $ 2.06 \pm 0.29 $ & $ 2.52 \pm 0.39 $   & $ 1.75 \pm 0.19 $ & $ 1.19 \pm 0.22 $ & $ 1.27 \pm 0.26 $     \\ 
  -0.75   &   $ 1.75 \pm 0.15 $ & $ 1.99 \pm 0.28 $ &        ---         & $ 2.19 \pm 0.17 $ & $ 1.90 \pm 0.18 $ & $ 1.89 \pm 0.20 $   & $ 2.07 \pm 0.18 $ & $ 1.50 \pm 0.20 $ & $ 1.69 \pm 0.20 $     \\ 
  -0.65   &   $ 1.32 \pm 0.12 $ & $ 1.94 \pm 0.20 $ &        ---         & $ 1.83 \pm 0.14 $ & $ 1.94 \pm 0.16 $ & $ 1.88 \pm 0.15 $   & $ 2.06 \pm 0.18 $ & $ 1.67 \pm 0.19 $ & $ 1.28 \pm 0.14 $     \\ 
  -0.55   &   $ 1.42 \pm 0.14 $ & $ 1.58 \pm 0.15 $ &        ---         & $ 1.42 \pm 0.12 $ & $ 1.78 \pm 0.14 $ & $ 1.71 \pm 0.14 $   & $ 2.13 \pm 0.19 $ & $ 1.50 \pm 0.18 $ & $ 1.58 \pm 0.17 $     \\ 
  -0.45   &   $ 1.12 \pm 0.10 $ & $ 1.75 \pm 0.15 $ &        ---         & $ 1.29 \pm 0.10 $ & $ 1.61 \pm 0.12 $ & $ 1.58 \pm 0.10 $   & $ 1.85 \pm 0.16 $ & $ 1.53 \pm 0.16 $ & $ 1.20 \pm 0.07 $     \\ 
  -0.35   &   $ 1.31 \pm 0.13 $ & $ 1.56 \pm 0.14 $ &        ---         & $ 1.25 \pm 0.11 $ & $ 1.61 \pm 0.11 $ & $ 1.68 \pm 0.13 $   & $ 1.72 \pm 0.15 $ & $ 1.37 \pm 0.15 $ & $ 1.90 \pm 0.18 $     \\ 
  -0.25   &   $ 1.20 \pm 0.13 $ & $ 1.58 \pm 0.14 $ &        ---         & $ 1.14 \pm 0.11 $ & $ 1.38 \pm 0.11 $ & $ 1.81 \pm 0.14 $   & $ 2.22 \pm 0.19 $ & $ 1.49 \pm 0.16 $ & $ 1.84 \pm 0.18 $     \\ 
  -0.15   &   $ 1.03 \pm 0.11 $ & $ 1.70 \pm 0.16 $ &        ---         & $ 1.22 \pm 0.11 $ & $ 1.46 \pm 0.12 $ & $ 1.57 \pm 0.12 $   & $ 2.02 \pm 0.18 $ & $ 1.57 \pm 0.15 $ & $ 1.58 \pm 0.15 $     \\ 
  -0.05   &   $ 1.44 \pm 0.19 $ & $ 1.40 \pm 0.14 $ &        ---         & $ 1.12 \pm 0.10 $ & $ 1.57 \pm 0.12 $ & $ 1.78 \pm 0.13 $   & $ 1.80 \pm 0.17 $ & $ 1.44 \pm 0.15 $ & $ 1.94 \pm 0.17 $     \\ 
   0.05   &   $ 1.19 \pm 0.17 $ & $ 1.55 \pm 0.15 $ & $ 1.27 \pm 0.12 $  & $ 0.97 \pm 0.09 $ & $ 1.46 \pm 0.11 $ & $ 1.59 \pm 0.12 $   & $ 1.56 \pm 0.16 $ & $ 2.01 \pm 0.18 $ & $ 1.72 \pm 0.16 $     \\ 
   0.15   &   $ 1.24 \pm 0.17 $ & $ 1.55 \pm 0.15 $ & $ 1.20 \pm 0.08 $  & $ 1.08 \pm 0.10 $ & $ 1.69 \pm 0.14 $ & $ 1.53 \pm 0.12 $   & $ 1.50 \pm 0.16 $ & $ 2.23 \pm 0.19 $ & $ 1.62 \pm 0.15 $     \\ 
   0.25   &   $ 1.58 \pm 0.24 $ & $ 1.96 \pm 0.18 $ & $ 1.13 \pm 0.10 $  & $ 1.09 \pm 0.15 $ & $ 1.77 \pm 0.14 $ & $ 1.37 \pm 0.11 $   & $ 1.83 \pm 0.18 $ & $ 1.90 \pm 0.14 $ & $ 1.85 \pm 0.17 $     \\ 
   0.35   &   $ 1.10 \pm 0.24 $ & $ 1.85 \pm 0.20 $ & $ 1.36 \pm 0.11 $  & $ 1.09 \pm 0.10 $ & $ 1.95 \pm 0.14 $ & $ 1.42 \pm 0.11 $   & $ 1.59 \pm 0.17 $ & $ 1.78 \pm 0.16 $ & $ 1.88 \pm 0.16 $     \\ 
   0.45   &   $ 1.52 \pm 0.32 $ & $ 1.74 \pm 0.19 $ & $ 1.24 \pm 0.11 $  & $ 1.38 \pm 0.12 $ & $ 1.67 \pm 0.14 $ & $ 1.79 \pm 0.13 $   & $ 1.46 \pm 0.19 $ & $ 1.79 \pm 0.17 $ & $ 1.88 \pm 0.17 $     \\ 
   0.55   &   $ 2.16 \pm 0.51 $ & $ 1.92 \pm 0.21 $ & $ 1.72 \pm 0.14 $  & $ 1.60 \pm 0.13 $ & $ 1.95 \pm 0.15 $ & $ 1.62 \pm 0.14 $   & $ 1.32 \pm 0.14 $ & $ 1.50 \pm 0.14 $ & $ 1.96 \pm 0.17 $     \\ 
   0.65   &   $ 3.39 \pm 0.88 $ & $ 1.66 \pm 0.18 $ & $ 1.89 \pm 0.15 $  & $ 1.74 \pm 0.14 $ & $ 1.98 \pm 0.16 $ & $ 1.75 \pm 0.17 $   & $ 1.21 \pm 0.15 $ & $ 1.49 \pm 0.15 $ & $ 1.74 \pm 0.16 $     \\ 
   0.75   &   $ 2.09 \pm 0.73 $ & $ 1.84 \pm 0.21 $ & $ 2.44 \pm 0.19 $  & $ 2.44 \pm 0.19 $ & $ 1.80 \pm 0.17 $ & $ 1.97 \pm 0.23 $   & $ 1.26 \pm 0.17 $ & $ 1.87 \pm 0.19 $ & $ 1.64 \pm 0.17 $     \\ 
   0.85   &   $ 2.64 \pm 1.56 $ & $ 1.96 \pm 0.35 $ & $ 2.58 \pm 0.32 $  & $ 2.90 \pm 0.29 $ & $ 2.04 \pm 0.29 $ & $ 2.04 \pm 0.44 $   & $ 1.10 \pm 0.22 $ & $ 1.88 \pm 0.21 $ & $ 2.15 \pm 0.28 $     \\ 
   0.95   &        ---          &       ---         & $ 3.70 \pm 0.86 $  & $ 4.33 \pm 0.59 $ & $ 1.86 \pm 0.53 $ & $ 3.13 \pm 1.66 $   & $ 1.04 \pm 0.40 $ & $ 1.74 \pm 0.23 $ & $ 2.44 \pm 0.37 $     \\ 
      \noalign{\smallskip}\hline
    \end{tabular}
\end{table*}

\begin{table*}
\caption{Cross sections in $\mathrm{\mu b/sr}$ for the reaction $\mathrm{pp\to pK^+\Lambda}$, $p_{beam} = 3059$ MeV/c, $\epsilon = 239$ MeV, fig.~\ref{fig:3059lambda}.} 
\label{tab:diffobs3059KpL}
         \begin{tabular}{@{}rccccccccc}
          \hline\noalign{\smallskip}
 cos$\,\theta$ & $\frac{d\sigma}{d\Omega}(\theta^*_\mathrm{p})$$\;$ & $\frac{d\sigma}{d\Omega}(\theta^*_\mathrm{K^+})$$\;$ & $\frac{d\sigma}{d\Omega}(\theta^*_\mathrm{\Lambda})$$\;$ & $\frac{d\sigma}{d\Omega}(\theta^{Rp\Lambda}_{bp})$$\;$ & $\frac{d\sigma}{d\Omega}(\theta^{RKp}_{bK})$$\;$ & $\frac{d\sigma}{d\Omega}(\theta^{RK\Lambda}_{bK})$$\;$ & $\frac{d\sigma}{d\Omega}(\theta^{Rp\Lambda}_{Kp})$$\;$ & $\frac{d\sigma}{d\Omega}(\theta^{RKp}_{\Lambda K})$$\;$ &  $\frac{d\sigma}{d\Omega}(\theta^{RK\Lambda}_{p\Lambda})$$\;$  \\ 
          \noalign{\smallskip}\hline\noalign{\smallskip}
  -0.95   &   $ 4.15 \pm 0.66 $ & $ 2.15 \pm 1.57 $ &        ---        & $ 4.52 \pm 0.98 $ & $ 2.32 \pm 0.51 $ & $ 3.04 \pm 0.70 $ & $ 2.63 \pm 0.20 $ & $ 2.10 \pm 0.50 $ & $ 1.45 \pm 0.36 $ 	\\ 
  -0.85   &   $ 2.72 \pm 0.23 $ & $ 2.07 \pm 0.42 $ &        ---        & $ 3.19 \pm 0.36 $ & $ 1.99 \pm 0.27 $ & $ 2.70 \pm 0.44 $ & $ 2.51 \pm 0.20 $ & $ 1.99 \pm 0.32 $ & $ 1.51 \pm 0.20 $ 	\\ 
  -0.75   &   $ 2.07 \pm 0.12 $ & $ 2.13 \pm 0.24 $ &        ---        & $ 2.37 \pm 0.15 $ & $ 1.74 \pm 0.15 $ & $ 2.35 \pm 0.25 $ & $ 2.54 \pm 0.18 $ & $ 1.47 \pm 0.13 $ & $ 1.38 \pm 0.12 $ 	\\ 
  -0.65   &   $ 1.68 \pm 0.10 $ & $ 1.91 \pm 0.17 $ &        ---        & $ 2.04 \pm 0.13 $ & $ 1.93 \pm 0.13 $ & $ 2.14 \pm 0.17 $ & $ 2.44 \pm 0.15 $ & $ 1.42 \pm 0.10 $ & $ 1.46 \pm 0.10 $ 	\\ 
  -0.55   &   $ 1.66 \pm 0.10 $ & $ 1.97 \pm 0.13 $ &        ---        & $ 1.76 \pm 0.10 $ & $ 1.77 \pm 0.12 $ & $ 2.08 \pm 0.14 $ & $ 2.43 \pm 0.15 $ & $ 1.56 \pm 0.11 $ & $ 1.49 \pm 0.09 $ 	\\ 
  -0.45   &   $ 1.54 \pm 0.10 $ & $ 2.02 \pm 0.12 $ &        ---        & $ 1.70 \pm 0.10 $ & $ 1.74 \pm 0.10 $ & $ 1.98 \pm 0.12 $ & $ 2.43 \pm 0.14 $ & $ 1.59 \pm 0.11 $ & $ 1.72 \pm 0.11 $ 	\\ 
  -0.35   &   $ 1.68 \pm 0.12 $ & $ 1.89 \pm 0.11 $ &        ---        & $ 1.38 \pm 0.08 $ & $ 1.74 \pm 0.10 $ & $ 1.92 \pm 0.11 $ & $ 2.46 \pm 0.15 $ & $ 1.76 \pm 0.11 $ & $ 1.76 \pm 0.12 $ 	\\ 
  -0.25   &   $ 1.73 \pm 0.11 $ & $ 1.89 \pm 0.12 $ &        ---        & $ 1.32 \pm 0.08 $ & $ 1.96 \pm 0.11 $ & $ 1.87 \pm 0.10 $ & $ 2.25 \pm 0.14 $ & $ 1.75 \pm 0.11 $ & $ 1.77 \pm 0.11 $ 	\\ 
  -0.15   &   $ 1.63 \pm 0.12 $ & $ 2.05 \pm 0.13 $ &        ---        & $ 1.23 \pm 0.07 $ & $ 1.87 \pm 0.10 $ & $ 1.81 \pm 0.10 $ & $ 2.22 \pm 0.14 $ & $ 2.10 \pm 0.13 $ & $ 1.91 \pm 0.12 $ 	\\ 
  -0.05   &   $ 1.64 \pm 0.16 $ & $ 1.95 \pm 0.12 $ &        ---        & $ 1.14 \pm 0.07 $ & $ 1.88 \pm 0.10 $ & $ 1.77 \pm 0.10 $ & $ 2.13 \pm 0.13 $ & $ 1.96 \pm 0.13 $ & $ 1.90 \pm 0.11 $ 	\\ 
   0.05   &   $ 1.49 \pm 0.15 $ & $ 2.16 \pm 0.14 $ & $ 1.16 \pm 0.07 $ & $ 1.20 \pm 0.07 $ & $ 2.01 \pm 0.11 $ & $ 1.75 \pm 0.10 $ & $ 2.02 \pm 0.12 $ & $ 2.45 \pm 0.18 $ & $ 2.05 \pm 0.12 $    \\ 
   0.15   &   $ 1.33 \pm 0.17 $ & $ 2.23 \pm 0.15 $ & $ 1.25 \pm 0.08 $ & $ 1.18 \pm 0.07 $ & $ 2.12 \pm 0.12 $ & $ 1.72 \pm 0.10 $ & $ 1.83 \pm 0.11 $ & $ 2.17 \pm 0.16 $ & $ 2.08 \pm 0.12 $    \\ 
   0.25   &   $ 1.45 \pm 0.23 $ & $ 2.40 \pm 0.15 $ & $ 1.22 \pm 0.08 $ & $ 1.18 \pm 0.07 $ & $ 2.02 \pm 0.11 $ & $ 1.72 \pm 0.10 $ & $ 1.69 \pm 0.11 $ & $ 2.32 \pm 0.14 $ & $ 2.13 \pm 0.13 $    \\ 
   0.35   &   $ 1.46 \pm 0.21 $ & $ 2.03 \pm 0.14 $ & $ 1.42 \pm 0.09 $ & $ 1.34 \pm 0.08 $ & $ 2.11 \pm 0.12 $ & $ 1.58 \pm 0.09 $ & $ 1.70 \pm 0.13 $ & $ 2.23 \pm 0.14 $ & $ 2.21 \pm 0.13 $    \\ 
   0.45   &   $ 1.32 \pm 0.20 $ & $ 2.11 \pm 0.15 $ & $ 1.62 \pm 0.10 $ & $ 1.41 \pm 0.09 $ & $ 1.97 \pm 0.11 $ & $ 1.67 \pm 0.12 $ & $ 1.53 \pm 0.11 $ & $ 2.23 \pm 0.13 $ & $ 2.19 \pm 0.14 $    \\ 
   0.55   &   $ 1.89 \pm 0.43 $ & $ 1.75 \pm 0.13 $ & $ 1.88 \pm 0.11 $ & $ 1.51 \pm 0.10 $ & $ 2.01 \pm 0.13 $ & $ 1.72 \pm 0.13 $ & $ 1.54 \pm 0.11 $ & $ 2.21 \pm 0.15 $ & $ 2.31 \pm 0.15 $    \\ 
   0.65   &   $ 2.27 \pm 0.53 $ & $ 1.61 \pm 0.18 $ & $ 2.10 \pm 0.12 $ & $ 1.79 \pm 0.11 $ & $ 1.97 \pm 0.15 $ & $ 1.85 \pm 0.15 $ & $ 1.27 \pm 0.10 $ & $ 2.21 \pm 0.15 $ & $ 2.37 \pm 0.15 $    \\ 
   0.75   &   $ 2.41 \pm 0.70 $ & $ 1.94 \pm 0.23 $ & $ 2.33 \pm 0.18 $ & $ 2.17 \pm 0.16 $ & $ 1.83 \pm 0.14 $ & $ 2.00 \pm 0.22 $ & $ 1.28 \pm 0.11 $ & $ 2.10 \pm 0.14 $ & $ 2.53 \pm 0.20 $    \\ 
   0.85   &   $ 1.73 \pm 0.76 $ & $ 2.02 \pm 0.39 $ & $ 3.02 \pm 0.45 $ & $ 3.00 \pm 0.28 $ & $ 1.81 \pm 0.24 $ & $ 2.09 \pm 0.39 $ & $ 1.32 \pm 0.17 $ & $ 2.22 \pm 0.19 $ & $ 2.58 \pm 0.28 $    \\ 
   0.95   &         ---       &        ---          & $ 3.91 \pm 1.25 $ & $ 4.32 \pm 0.56 $ & $ 1.97 \pm 0.59 $ & $ 2.41 \pm 1.14 $ & $ 0.99 \pm 0.17 $ & $ 2.30 \pm 0.25 $ & $ 2.95 \pm 0.44 $    \\ 
      \noalign{\smallskip}\hline
    \end{tabular}
\end{table*}

\begin{table*}[hhh]
\caption{Cross sections in $\mathrm{\mu b/sr}$ for the reaction $\mathrm{pp\to pK^+\Lambda}$, $p_{beam} = 3200$ MeV/c, $\epsilon = 284$ MeV, fig.~\ref{fig:3200lambda}.} 
\label{tab:diffobs3200KpL}
         \begin{tabular}{@{}rccccccccc}
          \hline\noalign{\smallskip}
 cos$\,\theta$ & $\frac{d\sigma}{d\Omega}(\theta^*_\mathrm{p})$$\;$ & $\frac{d\sigma}{d\Omega}(\theta^*_\mathrm{K^+})$$\;$ & $\frac{d\sigma}{d\Omega}(\theta^*_\mathrm{\Lambda})$$\;$ & $\frac{d\sigma}{d\Omega}(\theta^{Rp\Lambda}_{bp})$$\;$ & $\frac{d\sigma}{d\Omega}(\theta^{RKp}_{bK})$$\;$ & $\frac{d\sigma}{d\Omega}(\theta^{RK\Lambda}_{bK})$$\;$ & $\frac{d\sigma}{d\Omega}(\theta^{Rp\Lambda}_{Kp})$$\;$ & $\frac{d\sigma}{d\Omega}(\theta^{RKp}_{\Lambda K})$$\;$ &  $\frac{d\sigma}{d\Omega}(\theta^{RK\Lambda}_{p\Lambda})$$\;$  \\ 
          \noalign{\smallskip}\hline\noalign{\smallskip}
  -0.90   &  $ 4.41 \pm 0.52 $ & $ 2.45 \pm 1.31 $ &       ---          & $ 5.07 \pm 0.70 $ & $ 2.98 \pm 0.76 $ & $ 4.44 \pm 1.22 $ & $ 3.49 \pm 0.38 $ & $ 2.63 \pm 1.13 $ & $ 2.80 \pm 0.83 $    \\ 
  -0.70   &  $ 2.92 \pm 0.26 $ & $ 2.95 \pm 0.71 $ &       ---          & $ 2.89 \pm 0.28 $ & $ 2.85 \pm 0.45 $ & $ 3.03 \pm 0.51 $ & $ 3.37 \pm 0.32 $ & $ 2.34 \pm 0.49 $ & $ 1.84 \pm 0.34 $    \\ 
  -0.50   &  $ 2.11 \pm 0.21 $ & $ 2.96 \pm 0.36 $ &       ---          & $ 2.12 \pm 0.19 $ & $ 2.65 \pm 0.30 $ & $ 2.42 \pm 0.26 $ & $ 2.92 \pm 0.26 $ & $ 1.92 \pm 0.30 $ & $ 2.62 \pm 0.29 $    \\ 
  -0.30   &  $ 1.98 \pm 0.25 $ & $ 2.58 \pm 0.26 $ &       ---          & $ 1.90 \pm 0.17 $ & $ 2.18 \pm 0.18 $ & $ 2.18 \pm 0.20 $ & $ 2.77 \pm 0.29 $ & $ 2.31 \pm 0.26 $ & $ 1.72 \pm 0.20 $    \\ 
  -0.10   &  $ 2.59 \pm 0.40 $ & $ 2.92 \pm 0.29 $ &       ---          & $ 1.60 \pm 0.17 $ & $ 2.77 \pm 0.21 $ & $ 2.46 \pm 0.20 $ & $ 2.58 \pm 0.24 $ & $ 1.99 \pm 0.22 $ & $ 2.09 \pm 0.20 $    \\ 
   0.10   &  $ 1.83 \pm 0.42 $ & $ 2.76 \pm 0.29 $ & $ 1.45 \pm 0.35 $  & $ 1.71 \pm 0.17 $ & $ 2.46 \pm 0.20 $ & $ 1.98 \pm 0.18 $ & $ 2.15 \pm 0.22 $ & $ 2.65 \pm 0.27 $ & $ 2.26 \pm 0.24 $    \\ 
   0.30   &  $ 1.97 \pm 0.66 $ & $ 2.78 \pm 0.32 $ & $ 1.74 \pm 0.19 $  & $ 1.72 \pm 0.16 $ & $ 2.89 \pm 0.24 $ & $ 2.17 \pm 0.19 $ & $ 2.26 \pm 0.27 $ & $ 2.85 \pm 0.26 $ & $ 2.99 \pm 0.29 $    \\ 
   0.50   &  $ 2.16 \pm 0.99 $ & $ 2.43 \pm 0.32 $ & $ 2.15 \pm 0.20 $  & $ 1.82 \pm 0.17 $ & $ 2.78 \pm 0.26 $ & $ 2.45 \pm 0.28 $ & $ 2.01 \pm 0.26 $ & $ 2.57 \pm 0.23 $ & $ 2.82 \pm 0.32 $    \\ 
   0.70   &        ---         & $ 2.10 \pm 0.38 $ & $ 3.12 \pm 0.36 $  & $ 2.62 \pm 0.28 $ & $ 3.06 \pm 0.38 $ & $ 2.33 \pm 0.45 $ & $ 1.08 \pm 0.23 $ & $ 3.19 \pm 0.30 $ & $ 2.62 \pm 0.37 $    \\ 
   0.90   &        ---         &         ---       & $ 4.77 \pm 0.95 $  & $ 4.93 \pm 0.74 $ & $ 2.54 \pm 0.47 $ & $ 3.46 \pm 1.27 $ & $ 1.10 \pm 0.48 $ & $ 2.79 \pm 0.38 $ & $ 3.47 \pm 0.79 $     \\ 
      \noalign{\smallskip}\hline
    \end{tabular}
\end{table*}

\begin{table*}[hhh]
\caption{Cross sections in $\mathrm{\mu b/sr}$ for the reaction $\mathrm{pp\to pK^+\Sigma^0}$, $p_{beam} = 3059$ MeV/c, $\epsilon = 162$ MeV, fig.~\ref{fig:3059sigma}.} 
\label{tab:diffobs3059KpS0}
         \begin{tabular}{@{}rccccccccc}
          \hline\noalign{\smallskip}
  cos$\,\theta$ & $\frac{d\sigma}{d\Omega}(\theta^*_\mathrm{p})$$\;$ & $\frac{d\sigma}{d\Omega}(\theta^*_\mathrm{K^+})$$\;$ & $\frac{d\sigma}{d\Omega}(\theta^*_\mathrm{\Sigma^0})$$\;$ & $\frac{d\sigma}{d\Omega}(\theta^{Rp\Sigma^0}_{bp})$$\;$ & $\frac{d\sigma}{d\Omega}(\theta^{RKp}_{bK})$$\;$ & $\frac{d\sigma}{d\Omega}(\theta^{RK\Sigma^0}_{bK})$$\;$ & $\frac{d\sigma}{d\Omega}(\theta^{Rp\Sigma^0}_{Kp})$$\;$ & $\frac{d\sigma}{d\Omega}(\theta^{RKp}_{\Sigma^0 K})$$\;$ &  $\frac{d\sigma}{d\Omega}(\theta^{RK\Sigma^0}_{p\Sigma^0})$$\;$  \\ 
         \noalign{\smallskip}\hline\noalign{\smallskip}
  -0.90   &  $ 0.26 \pm 0.05 $ & $ 0.34 \pm 0.11 $ &       ---         &  $ 0.33 \pm 0.07 $ & $ 0.33 \pm 0.10 $ & $ 0.40 \pm 0.12 $    & $ 0.36 \pm 0.07 $ & $ 0.35 \pm 0.12 $ & $ 0.16 \pm 0.05 $    \\ 
  -0.70   &  $ 0.31 \pm 0.05 $ & $ 0.30 \pm 0.06 $ &       ---         &  $ 0.27 \pm 0.05 $ & $ 0.35 \pm 0.07 $ & $ 0.28 \pm 0.06 $    & $ 0.32 \pm 0.06 $ & $ 0.34 \pm 0.07 $ & $ 0.18 \pm 0.04 $    \\ 
  -0.50   &  $ 0.29 \pm 0.05 $ & $ 0.26 \pm 0.04 $ &       ---         &  $ 0.27 \pm 0.04 $ & $ 0.31 \pm 0.05 $ & $ 0.27 \pm 0.05 $    & $ 0.27 \pm 0.05 $ & $ 0.25 \pm 0.05 $ & $ 0.26 \pm 0.05 $    \\ 
  -0.30   &  $ 0.23 \pm 0.04 $ & $ 0.26 \pm 0.04 $ &       ---         &  $ 0.27 \pm 0.04 $ & $ 0.22 \pm 0.04 $ & $ 0.28 \pm 0.04 $    & $ 0.28 \pm 0.05 $ & $ 0.27 \pm 0.05 $ & $ 0.24 \pm 0.04 $    \\ 
  -0.10   &  $ 0.23 \pm 0.04 $ & $ 0.23 \pm 0.04 $ &       ---         &  $ 0.25 \pm 0.04 $ & $ 0.21 \pm 0.03 $ & $ 0.25 \pm 0.04 $    & $ 0.28 \pm 0.05 $ & $ 0.29 \pm 0.05 $ & $ 0.28 \pm 0.05 $    \\ 
   0.10   &  $ 0.23 \pm 0.05 $ & $ 0.18 \pm 0.03 $ & $ 0.30 \pm 0.05 $ &  $ 0.26 \pm 0.04 $ & $ 0.20 \pm 0.03 $ & $ 0.27 \pm 0.04 $    & $ 0.25 \pm 0.04 $ & $ 0.29 \pm 0.05 $ & $ 0.29 \pm 0.05 $    \\ 
   0.30   &  $ 0.29 \pm 0.07 $ & $ 0.26 \pm 0.05 $ & $ 0.28 \pm 0.05 $ &  $ 0.26 \pm 0.04 $ & $ 0.21 \pm 0.03 $ & $ 0.27 \pm 0.04 $    & $ 0.25 \pm 0.04 $ & $ 0.25 \pm 0.04 $ & $ 0.30 \pm 0.05 $    \\ 
   0.50   &  $ 0.27 \pm 0.09 $ & $ 0.31 \pm 0.06 $ & $ 0.28 \pm 0.05 $ &  $ 0.30 \pm 0.05 $ & $ 0.26 \pm 0.04 $ & $ 0.27 \pm 0.04 $    & $ 0.28 \pm 0.05 $ & $ 0.29 \pm 0.05 $ & $ 0.31 \pm 0.06 $    \\ 
   0.70   &  $ 0.40 \pm 0.20 $ & $ 0.32 \pm 0.08 $ & $ 0.28 \pm 0.05 $ &  $ 0.30 \pm 0.06 $ & $ 0.32 \pm 0.07 $ & $ 0.27 \pm 0.06 $    & $ 0.21 \pm 0.05 $ & $ 0.23 \pm 0.04 $ & $ 0.25 \pm 0.05 $    \\ 
   0.90   &        ---        &  $ 0.35 \pm 0.17 $ & $ 0.34 \pm 0.10 $ &  $ 0.25 \pm 0.06 $ & $ 0.34 \pm 0.12 $ & $ 0.33 \pm 0.12 $    & $ 0.19 \pm 0.06 $ & $ 0.32 \pm 0.06 $ & $ 0.36 \pm 0.08 $    \\ 
      \noalign{\smallskip}\hline
    \end{tabular}
\end{table*}

\end{document}